\documentclass{article}

\PassOptionsToPackage{numbers, compress}{natbib}
 \usepackage[preprint]{neurips_2020}

\usepackage[utf8]{inputenc} % allow utf-8 input
\usepackage[T1]{fontenc}    % use 8-bit T1 fonts
\usepackage[hyphens]{url}            % simple URL typesetting
\usepackage[breaklinks]{hyperref}       % hyperlinks
\usepackage{booktabs}       % professional-quality tables
\usepackage{amsfonts}       % blackboard math symbols
\usepackage{nicefrac}       % compact symbols for 1/2, etc.
\usepackage{microtype}      % microtypography
\usepackage{graphicx}
\usepackage{bm}
\usepackage{mathtools}
\usepackage[linesnumbered,ruled]{algorithm2e}
\usepackage{caption}
\usepackage{longtable}
\usepackage{multirow}
\usepackage[resetlabels]{multibib}
\usepackage{authblk}

\newcites{S}{Supplementary references}

\newcommand{\nocontentsline}[3]{}
\newcommand{\tocless}[2]{\bgroup\let\addcontentsline=\nocontentsline#1{#2}\egroup}

\title{Parametric Copula-GP model for analyzing multidimensional neuronal and behavioral relationships}

\author[1]{\textbf{Nina Kudryashova}}
\author[2]{\textbf{Theoklitos Amvrosiadis}}
\author[2]{\textbf{Nathalie Dupuy}}
\author[2,3]{\textbf{Nathalie Rochefort}}
\author[1]{\textbf{Arno~Onken}}
\affil[1]{School of Informatics, University of Edinburgh}
\affil[2]{Centre for Discovery Brain Sciences, University of Edinburgh}
\affil[3]{Simons Initiative for the Developing Brain, University of Edinburgh}
\affil[ ]{~}
\affil[ ]{\texttt{nkudryas@inf.ed.ac.uk, t.amvrosiadis@ed.ac.uk, nathalie.dupuy@ed.ac.uk, n.rochefort@ed.ac.uk, aonken@inf.ed.ac.uk}}
\begin{document}

\maketitle

\begin{abstract}
    One of the main challenges in current systems neuroscience is the analysis of high-dimensional neuronal and behavioral data that are characterized by different statistics and timescales of the recorded variables. We propose a parametric copula model which separates the statistics of the individual variables from their dependence structure, and escapes the curse of dimensionality by using vine copula constructions. We use a Bayesian framework with Gaussian Process (GP) priors over copula parameters, conditioned on a continuous task-related variable. We validate the model on synthetic data and compare its performance in estimating mutual information against the commonly used non-parametric algorithms. 
    %(KSG, bias-improved KSG and MINE) %fits to the page so well without it!!!
    Our model provides accurate information estimates when the dependencies in the data match the parametric copulas used in our framework. When the exact density estimation with a parametric model is not possible, our Copula-GP model is still able to provide reasonable information estimates, close to the ground truth and comparable to those obtained with a neural network estimator. Finally, we apply our framework to real neuronal and behavioral recordings obtained in awake mice. We demonstrate the ability of our framework to 
    1)~produce accurate and interpretable bivariate models for the analysis of inter-neuronal noise correlations or behavioral modulations;
    2)~expand to more than 100 dimensions and measure information content in the whole-population statistics. These results demonstrate that the Copula-GP framework is particularly useful for the analysis of complex multidimensional relationships between neuronal, sensory and behavioral data.

\end{abstract}

%defs
\def\GPU{RTX 2080Ti}
\def\WAIC{\mathrm{WAIC}}
\def\lppd{\mathrm{lppd}}
\def\MI{\mathrm{MI}}
\def\bY{\mathbf{Y}}
\def\bU{\mathbf{U}}
\def\bF{\mathbf{F}}

\tocless\section{Introduction}

Recent advances in imaging and recording techniques have enabled monitoring the activity of hundreds to several thousands of neurons simultaneously~\cite{jun2017fully,helmchen2009two,dombeck2007imaging}.
These recordings can be made in awake animals engaged in specifically designed tasks or natural behavior~\cite{stringer2019spontaneous,pakan2018impact,pakan2018action}, which further augments these already large datasets with a variety of behavioral variables.
% These newly designed tools must be difference in statistics and time-scales of the recorded variables, which ... the activity can be recorded in different ways.
% These neural and behavioral variables operate at different timescales and exhibit different statistics. 
% The rich, large datasets that emerge provide new insights into brain function, but their complexity and high dimensionality necessitates the development of novel analytical approaches (cunnigham,yu ref). Neural and behavioral variables operate at different timescales and exhibit different statistics. 
% Developments in neural recording techniques has enabled acquisition of hundreds to thousands of neuronal signals simultaneously~\cite{jun2017fully, dombeck2007imaging}. 
% These recordings can be made in awake behaving animals, which further enriches these already large datasets with a variety of behavioral variables~\cite{stringer2019spontaneous,pakan2018impact}. 
% The analysis of these datasets becomes challenging due to their high dimensionality and difference in statistics and time-scales of the recorded variables.
% There is no single recording technique for neural activity either, but rather a number of different methods, such as neuropixels probes~\cite{jun2017fully} that measure instantaneous local field potentials or 2-photon calcium imaging~\cite{dombeck2007imaging} that records slow traces of the neuronal activity.
These complex high dimensional datasets necessitate the development of novel analytical approaches~\cite{staude2010cubic,brown2004multiple,Saxena2019,Stevenson2011} to address two central questions of systems and behavioral neuroscience: how do populations of neurons encode information? And how does this neuronal activity correspond to the observed behavior? 
In~machine learning terms, both of these questions translate into understanding the high-dimensional multivariate dependencies between the recorded variables~\cite{ince2010information,shamir2004nonlinear,kohn2016correlations,shimazaki2012state}.
% stringer2019spontaneous? 

There are two major methods suitable for recording the activity of large populations of neurons from behaving animals:
the multi-electrode probes that provide milliseconds precision for recordings of electrical activity~\cite{jun2017fully}, and calcium imaging methods~\cite{helmchen2009two,dombeck2007imaging,grienberger2015dendritic} that use changes in intracellular calcium concentration as a proxy for neuronal spiking activity at a lower (tens of milliseconds) temporal precision.
As a result, the recorded neuronal and behavioral variables may operate at different timescales and exhibit different statistics, which further complicates the statistical analysis.% of these datasets.

The natural approach to modeling statistical dependencies between the variables with drastically different statistics is based on \emph{copulas}, which separate marginal statistics from the dependence structure~\cite{joe2014dependence}. 
For this reason, copula models are particularly effective for mutual information estimation~\cite{jenison2004shape,calsaverini2009information}.
They can also escape the `curse of dimensionality' by factorising the multi-dimensional dependence into pair-copula constructions called \emph{vines}~\cite{aas2009pair,czado2010pair}.
Copula models have been successfully applied to spiking activity~\cite{berkes2009characterizing, onken2009analyzing,hu2015copula,shahbaba2014a}, 2-photon calcium recordings~\cite{safaai2019} and multi-modal neuronal datasets~\cite{onken2016nips}.
% ...\cite{li2011functional} %firing rate
However, these models assumed that the dependence between variables was static, % independent of the time or the inputs,
whereas in neuronal recordings it may be dynamic or modulated by behavioral context~\cite{doiron2016mechanics,shimazaki2012state}.
% The relationship between neurons and behavioral variables may be time- or context-dependent.
Therefore, it might be helpful to explicitly model the continuous time- or context-dependent changes in the relationships between variables, which reflect changes in an underlying computation.% for certain datasets. 
% Considering changes in the dependence can be essential for analysis of the timeseries or recordings aligned with some other variable (e.g. spatial location of an animal), which designates the phase of the task. 
% In such recordings, the dependence between neurons and behavioral variables may vary during the task,
% reflecting the changes in an underlying computation that takes place in each phase. 

Here, we extend a copula-based approach by adding explicit conditional dependence to the parameters of the copula model, approximating these latent dependencies with Gaussian Processes (GP).
It was previously shown that %outside of neuroscience,
such a combination of parametric copula models with GP priors outperforms static copula models~\cite{lopez2013gaussian} and even dynamic copula models on many real-world datasets, including weather forecasts, geological data or stock market data~\cite{hernandez2013nips}. 
%We show that neuronal datasets also involve changes in dependency structure. 
Yet, this method has never been applied to neuronal recordings before.

%here we need to highlight the novelty
In this work, we improve the scalability of the method by using stochastic variational inference. We also increase the complexity of the copula models in order to adequately describe the complex dependencies commonly observed in neuronal data. In particular, we use mixtures of parametric copula models to account for changes in tail dependencies.
We develop model selection algorithms, based on the fully-Bayesian Watanabe–Akaike information criterion (WAIC). 
Finally and most importantly, we demonstrate that our model is suitable for estimating mutual information.
It performs especially well when the parametric model can closely approximate the target distribution.
When it is not the case, our copula mixture model demonstrates sufficient flexibility and provides close information estimates, comparable to the best state-of-the-art non-parametric information estimators.
%on both synthetic and real multidimensional data. %very strong claim, lets see if I can provide enough evidence

We first introduce the copula mixture models and propose model selection algorithms (Sec.~2). %~\ref{sec:model} 
We then validate our model on synthetic data and compare its performance against other commonly used information estimators (Sec.~3). %~\ref{sec:artificial} 
Next, we demonstrate the utility of the method on real neuronal and behavioral data (Sec.~4). %~\ref{sec:real} 
We show that our Copula-GP method can produce interpretable bivariate models that emphasize the qualitative changes in tail dependencies and
%We demonstrate that our model can 
estimate mutual information that exposes the structure of the task without providing any explicit cues to the model. 
Finally, we apply the vine Copula-GP model to measure information content in the whole dataset with 5 behavioral variables and more than 100 neurons. 

\tocless\section{Parametric copula mixtures with Gaussian process priors}
\label{sec:model}

Our model is based on copulas: multivariate distributions with uniform marginals.
Sklar's theorem~\cite{sklar1959fonctions} states that any multivariate joint distribution can be written in terms of univariate marginal distribution functions $p(Y_i)$ and
a unique copula which characterizes the dependence structure:
$p(Y_1,\ldots,Y_N) = c(F_1(Y_1) \ldots F_N(Y_N)) \times \prod_{i=1}^N p(Y_i)$.
Here, $F_i(\cdot)$ are the marginal cumulative distribution functions~(CDF) and, as a result, each $F_i(Y_i)$ is uniformly distributed on [0,1].

%In this paper, we use semi-parametric approach: non-parametric marginals and parametric copulas.
For high dimensional datasets (high $\mathrm{dim}\bY$), maximum likelihood estimation for copula parameters may become computationally challenging. The two-stage inference for margins (IFM) training scheme is typically used in this case~\cite{joe2005}. First, univariate marginals are estimated and used to map the data onto a multidimensional unit cube. Second, the parameters of the copula model are inferred. %assuming that the marginals of the transformed data are uniform and fixed. 
%problem here. Joe2005 paper only talks about MLE, how to justify non-parametric marginal models?
% This algorithm was shown to be asymptotically equivalent to the maximum likelihood estimation~\cite{joe2005} (in a limit of sample size $\rightarrow \infty$).

\paragraph{Conditional copulas}
Following the approach by \citet{hernandez2013nips}, we are using Gaussian Processes~(GP) to model the conditional dependencies of copula parameters:
\begin{equation}
    p(\bY|X) = c\Big(F_1(Y_1|X), \ldots, F_N(Y_N|X) 
    \Big| X\Big) \times \left[\prod_{i=1}^N p(Y_i|X)\right].
    \label{eq:copula}
\end{equation}
In the most general case, the marginal PDFs $p(Y_i|X)$ and CDFs $F_i(Y_i|X)$ and the copula $c(\ldots|X)$ itself can all be conditioned on $X$. 
In our framework, $X$ is assumed to be one-dimensional.
A Gaussian Process is ideally suited for copula parametrization, as it provides an estimate of the uncertainty in model parameters, which we utilize in our model selection process (Sec.~2.1). %~\ref{sec:select}
%and in estimation of information measures. 

% If the dependence is expected to vary with respect to a certain variable ($X$), one may consider a conditional copula~\cite{hernandez2013nips}:
% \begin{equation}
%     p(Y_1,\ldots,Y_N|X) = c(F_1^X(Y_1|X) \ldots F_N^X(Y_N|X)) \times \prod_{i=1}^N p(Y_i|X)
% \end{equation}
% The ... transformed data reveals the dependency between the variables, which can also be dependent on $X$. 
%However, the same method can be extended to higher dimensions of $X$. %DEL?

\paragraph{Conditional marginals}

In order to estimate marginal CDFs $F(Y_i|X)$, we use the non-parametric fastKDE~\cite{obrien2016} algorithm, which allows for direct estimation of the conditional distributions. The conditional distribution is then used to map the data onto a unit hypercube using the probability integral transform:
$F(Y_i|X) \rightarrow U_i \sim U_{[0,1]}$, such that $U_i$ is uniformly distributed for any $X$.

\begin{figure}[t]
    \centering
    \includegraphics[width=\textwidth]{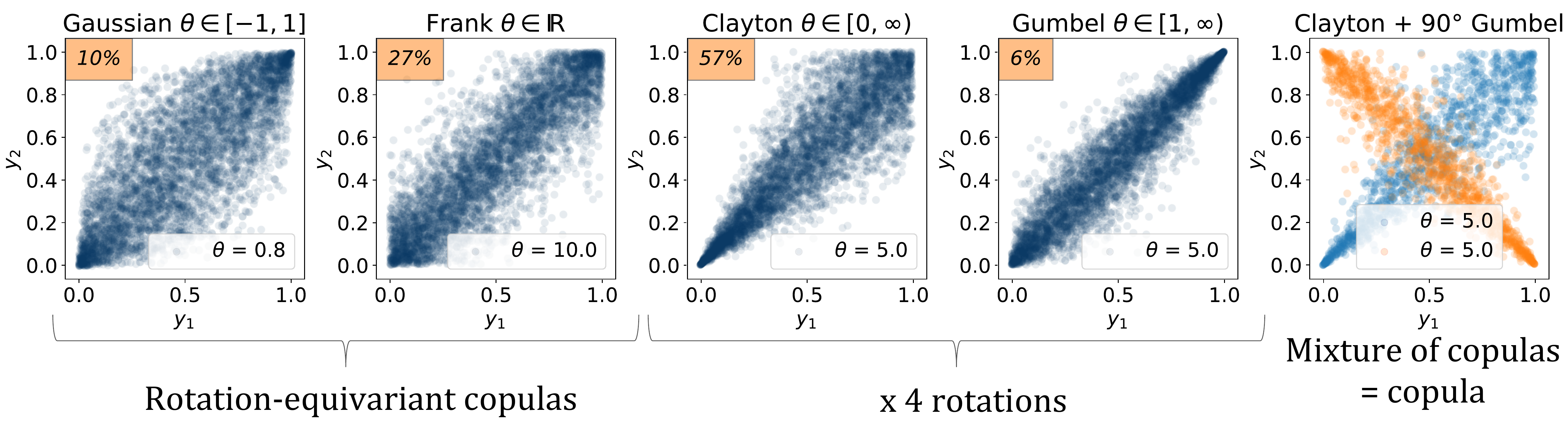}
    \caption{Copula families used in the mixture models in our framework. 
    The percentage in the upper-left corner shows how often each of the families was selected to be used in a copula mixture for pairwise relationships in the real neuronal data from~\citet{pakan2018impact} (see Sec.~4). %~\ref{sec:real}
    % Note, that considering the rotation, there are 10 copula elements in total. %DEL?
    }
    \label{fig:fam}
\end{figure}

\paragraph{Bivariate copula families}
We use 4~copula families as the building blocks for our copula models: Gaussian, Frank, Clayton and Gumbel copulas~(Figure~\ref{fig:fam}). All of these families have a single parameter, corresponding to the rank correlation~(Table~\ref{tab:families}).
We also use rotated variants (90$^\circ$, 180$^\circ$, 270$^\circ$) of Clayton and Gumbel copula families in order to express upper tail dependencies and negative correlation.

\begin{table}[h]
  \caption{Bivariate copula families and their GPLink functions}
  \label{tab:families}
  \centering
  \begin{tabular}{llll}
    \toprule
    % & \multicolumn{2}{c}{Domain}                   \\
    % \cmidrule(r){2-3}
    Copula       & Domain            & $\mathrm{GPLink}(f): \mathbb R \rightarrow \mathop{\mathrm{dom}}(c_j)$ \\
    \midrule
    Independence & --           & --     \\
    Gaussian     & [-1,1]       & $\mathrm{Erf}(f/1.4)$ \\
    Frank        & (-$\infty$, $\infty$) & $0.3 \cdot f + \mathrm{sign}(f)\cdot(0.3\cdot f)^2$  \\
    Clayton      & [0,$\infty$) & $\mathrm{Exp}(0.3\cdot f)$  \\
    Gumbel       & [1,$\infty$) & $1 + \mathrm{Exp}(0.3\cdot f)$  \\
    \bottomrule
  \end{tabular}
\end{table}

Since we are primarily focused on the analysis of neuronal data, we have first visualized the dependencies in calcium signal recordings after a probability integral transform, yielding empirical conditional copulas. As a distinct feature in neuronal datasets, we observed changes in tail dependencies with regard to the conditioning variable.
Since none of the aforementioned families alone could describe such conditional dependency, we combined multiple copulas into a linear \emph{mixture model} (which is also a copula~\cite{nelsen2007introduction}):
%Given a dataset $\left\{X_i,\vec Y_i\right\}_N$, where $X$ 
\begin{equation}
    c\left(\bU \middle| X\right) = \sum_{j=1}^{K} \phi_j(X) c_j(\bU; \theta_j(X)),
    \label{eq:mix}
\end{equation}
where $K$ is the number of elements, $\phi_j(X)$ is the concentration of the $j$th copula in a mixture, $c_j$ is the pdf of the $j$th copula, and $\theta_j$ is its parameter.

Each of the copula families includes the Independence copula as a special case. To resolve this overcompleteness, we add the Independence copula as a separate model with zero parameters (Table~\ref{tab:families}). For independent variables $\bY_{ind}$, the Independence model will be preferred over the other models in our model selection algorithm (Sec.~2.1), since it has the smallest number of parameters. %~\ref{sec:select}

\paragraph{Gaussian Process priors}

We parametrize the mixture model (\ref{eq:mix}) with the independent latent GPs:
$\mathbf{f} \sim \mathcal{N}(\mu\times\mathbf{1},K_\lambda(X,X))$.
For each copula family, we constructed GPLink functions (Table~\ref{tab:families}) that map the GP variable onto the copula parameter domain:
$ %\qquad \theta_j \in \mathrm{A(Copula_j)}
\theta_j = \mathrm{GPlink}_{c_j}(f_j), \mathbb R \rightarrow \mathop{\mathrm{dom}}(c_j).$
Next, we also use GP to parametrize concentrations $\phi_j(X)$, which are defined on a simplex ($\sum \phi = 1$):
% We use the following parametrisation:
    $$\phi_j = (1-t_j)\prod_{m=1}^{j-1} t_m,
    \quad t_m = \Phi\left(\widetilde f_m + \Phi^{-1} \left(\frac{M-m-1}{M-m}\right)\right),
    \quad t_M = 0,
    $$
where $\Phi$ is a CDF of a standard normal distribution and $\mathbf{\widetilde f_m} \sim \mathcal{N}(\widetilde{\mu}_m\times\mathbf{1},\widetilde{K}_{\widetilde\lambda_m}(X,X))$. 
This parametrization ensures that when all GP variables $\widetilde f_m = 0$, all of the concentrations $\phi_j$ are equal to $1/M$.
% $z \sim \mathrm{Categorical}(\phi),\\$
We use the RBF~kernel $K_\lambda(X,X)$ with bandwidth parameter $\lambda$.
Therefore, the whole mixture model with $M$ copula elements requires $[2M-1]$ hyperparameters: $\{\lambda\}_M$ for $\bm\theta$ and $\{\widetilde \lambda\}_{M-1}$ for $\bm\phi$. 
% $\{\lambda\}_K$, $\{\widetilde \lambda\}_{K-1}$ are GP kernel hyperparameters. 

\paragraph{Approximate Inference}

Since our model has latent variables with GP priors and intractable posterior distribution, the direct maximum likelihood Type-II estimation is not possible and an approximate inference is needed. Such inference problem with copula models has previously been solved with the %iterative
expectation propagation (EP) algorithm~\cite{hernandez2013nips}. Considering the recent developments in high-performance parallel computing and stochastic optimization algorithms, we chose to use \emph{stochastic variational inference} (SVI) instead. 
In order to scale the SVI to a large number of inputs, we use Kernel Interpolation for Scalable Structured Gaussian Processes (KISS-GP)~\cite{wilson2015kernel}. 
%This method works especially well on structured data (such as multi-trial time series with measurements taken at fixed intervals), since it places the inducing points on a fixed regular grid. %EXT?
For efficient implementation of these methods on 
%graphic processing units %EXT?
GPU, we use the PyTorch~\cite{paszke2017automatic} and GPyTorch libraries~\cite{gardner2018gpytorch}.

\tocless\subsection{Bayesian Model selection}
\label{sec:select}

We use the Watanabe–Akaike information criterion (WAIC~\cite{watanabe2013widely}) for model selection. WAIC is a fully Bayesian approach to estimating the Akaike information criterion (AIC) (see Eq.~31 in the original paper~\cite{watanabe2013widely}). The main advantage of the method is that it avoids the empirical estimation of the effective number of parameters, which is often used for approximation of the out-of-sample bias. It starts with the estimation of the log pointwise posterior predictive density (lppd)~\cite{gelman2014understanding}:
$$\widehat{\lppd} = \sum_{i=1}^{N}\log\left(
\frac1S \sum_{s=1}^{S} p(y_i|\theta^s)
\right),
\qquad
p_{\WAIC} = \sum_{i=1}^{N} V_{s=1}^{S} 
\Big(\log p(y_i|\theta^s)\Big),$$
where $\{\theta^s\}_S$ is a draw from a posterior distribution, which must be large enough to represent the posterior. Next, the $p_{\WAIC}$ approximates the bias correction, where $V_{s=1}^{S}$ represents sample variance.
%This approximation was shown to be the closest one to the leave-one-out cross-validation estimate of the bias~\cite{watanabe2010asymptotic}.
Therefore, the bias-corrected estimate of the log pointwise posterior predictive density is given by:
$e\widehat{\lppd}_\WAIC = \lppd - p_{\WAIC} = -N\cdot{\WAIC_{original}}.$

In the model selection process, we aim to choose the model with the lowest $\WAIC$. 
Since our copula probability densities are continuous, their values can exceed 1 and the resulting $\WAIC$ is typically negative. Zero $\WAIC$ corresponds to the Independence model ($\mathrm{pdf} = 1$ on the whole unit square). 
%We also set up a tolerance ($\WAIC_{tol} = 0.005$), and models with $\WAIC \in [-\WAIC_{tol},\WAIC_{tol}]$ are considered indistinguishable from the independence model.

Since the total number of combinations of 10~copula elements (Fig.~\ref{fig:fam}, considering rotations) is large, exhaustive search for the optimal model is not feasible.
In our framework, we propose two model algorithms for constructing close-to-optimal copula mixtures:
\emph{greedy} and \emph{heuristic} (see Supplemental Material for details). 
The greedy algorithm is universal and can be used with any other copula families without adjustment, while the heuristic algorithm is fine-tuned to the specific copula families used in this paper (Fig.~\ref{fig:fam}).
Both model selection algorithms were able to select the correct 1- and 2-component model on simulated data and at least find a close approximation (within $\WAIC_{tol} = 0.005$) for more complex models (see validation of model selection in Supplemental Material). 

\tocless\subsection{Entropy and mutual information}

Our framework provides tools for efficient sampling from the conditional distribution and for calculating the probability density $p(\bY|X)$. 
Therefore, for each $X=x$ the entropy $H(\bY|X=x)$ can be estimated using Monte Carlo (MC) integration:
\begin{equation}
H(\bY|X=x)= -\mathop{\mathbb{E}}_{p(\bY|X=x)} \log p(\bY|X=x).
%= \sum_{i=1}^N \H(Y_i|X=x) -\mathop{\mathbb{E}}_{c(U^x|X=x)} \log c(U^x|X=x),
\label{eq:Hx}
\end{equation}
$p(\bY|X=x)$ factorizes into the conditional copula density and marginal densities~(\ref{eq:copula}), hence for each $x$ the entropy also factorizes~\cite{jenison2004shape} as 
$H(\bY|X=x) = \sum H(Y_i|X=x) + H_c(\bU^X|X=x)$, 
where $\bU^X = \bF(\bY|X)$.
The conditional entropy can be integrated as
$H(\bY|X) = \sum_{i=1}^N H(Y_i|X) + \int H_c(\bU^X|X=x)p(x)dx,$
separating the entropy of the marginals $\{Y_i\}_N$ from the copula entropy.

Now, $I(X,\bY) = I(X,\mathbf{G(Y)})$ if $\mathbf{G(Y)}$ is 1)~a~homeomorphism, 2)~independent of~$X$~\cite{kraskov2004estimating}. 
If marginal statistics are independent of $X$, then the probability integral transform $\bU = \bF(\bY)$ satisfies both requirements, and $I(X,\bY) = I(X,\bU)$.
Then, in order to calculate the mutual information $I(X,\bU) \coloneqq H(\bU) - H(\bU|X)$, we must also rewrite it using only the conditional distribution $p(\bU|X)$, which is modelled with our conditional Copula-GP model.
This can be done as follows:
\begin{equation}
I(X,\bU) = H(\bU) - \int H(\bU|X=x)p(x)dx =
\mathop{\mathbb{E}}_{p(\bU,X)} \log p(\bU|X) - \mathop{\mathbb{E}}_{p(\bU)} \log \mathop{\mathbb{E}}_{p(X)} p(\bU|X).
\label{eq:MI}
\end{equation}
The last term in (\ref{eq:MI}) involves nested integration, which is computationally difficult and does not scale well with $N = \mathrm{dim} \bU$.
%Also, one can see that if we rewrite $I(X,\bY)$ the same way as (\ref{eq:MI}), this nested integration term forbids the factorisation as in (\ref{eq:H}) (see proof in Supplementary Material).
Therefore, we propose an alternative way of estimating $I(X,\bY)$, which avoids double integration and allows us to use the marginals conditioned on $X$ ($\bU^X = \bF(\bY|X)$), providing a better estimate of $H(\bY|X)$.
We can use two separate copula models, one for estimating $p(\bY)$ and calculating $H(\bY)$, and another one for estimating $p(\bY|X)$ and calculating $H(\bY|X)$:
\begin{equation}
    I(X,\bY) = \sum_{i=1}^N I(X,Y_i) + 
    H_{c} {(u_1,\ldots,u_N)} - \int H_c(u^x_1,\ldots,u^x_N|s=x)p(x)dx,
    \label{eq:MI_alt}
\end{equation}
where both entropy terms are estimated with MC (\ref{eq:Hx}). Here we only integrate over the unit cube $[0,1]^N$ and then $\mathrm{dom}\,X$, whereas (\ref{eq:MI}) required integration over $[0,1]^N \times \mathrm{dom}\,X$.

The performance of both (\ref{eq:MI}) and (\ref{eq:MI_alt}) critically depends on the approximation of the dependence structure, i.e. how well the parametric copula approximates the true copula probability density. 
If the joint distribution $p(Y_1\ldots Y_N)$ has a complex dependence structure, as we will see in synthetic examples, then the mixture of parametric copulas may provide a poor approximation of $p(\bY)$ and overestimate $H_{c} {(u_1,\ldots,u_N)}$, thereby overestimating $I(X,\bY)$. 
The direct integration (\ref{eq:MI}), on the other hand, typically underestimates the $I(X,\bY)$ due to imperfect approximation of $p(\bY|X)$, but it is only applicable if the marginals can be considered independent of $X$.

We further refer to the direct integration approach (\ref{eq:MI}) as "Copula-GP integrated" and to the alternative approach (\ref{eq:MI_alt}) as "Copula-GP estimated" and assess both of them on synthetic and real data.

\tocless\subsection{Copula vine constructions}
\label{sec:vine}

High-dimensional copulas can be constructed from bivariate copulas by organizing them into hierarchical structures called \emph{copula vines}~\cite{aas2009pair}.
% There are many possible decompositions based on different assumptions about conditional independence of specific elements in a model, which can be classified using graphical models called regular vines~\cite{}. 
In this paper, we focus on the \emph{canonical vine} or \emph{C-vine}, which factorizes the high-dimensional copula probability density function as follows:
\begin{equation}
    c(\bU) = \left[\prod_{i=2}^N c_{1i}(U_1,U_i)\right] \times \left[
    \prod_{i=2}^N \prod_{j=i+1}^N c_{ij|\{k\}_{k<i}}\left (F(U_i|\{U_k\}_{k<i}),F(U_j|\{U_k\}_{k<i})\right) \right]
    \label{eq:Cvine}
\end{equation}
where $\{k\}_{k<i} = 1, \ldots, i-1$, $\{U_k\}_{k<i} = U_1, \ldots, U_{i-1}$ and $F(U_i|\{U_k\}_{k<i})$ is a conditional CDF.
Note, that all of the copulas in~(\ref{eq:Cvine}) can also be conditioned on $X$ via Copula-GP model.
% We use heuristic element ordering based on the sum of absolute values of Kendall's $\tau$ of a given element with all of the other elements. Thus,
We choose the first variable $U_1$ to be the one with the highest rank correlation with the rest (sum of absolute values of pairwise Kendall's $\tau$), and condition all variables on the first one.
We repeat the procedure until no variable is left.
It was shown by~\citet{czado2012maximum} that this ordering facilitates C-vine modeling.
% The C-vine was also shown to be a good choice for neuronal datasets~\cite{onken2009analyzing}, as they often include some proxy of a neuronal population activity as an outstanding variable, strongly correlated with the rest.
\paragraph{Code availability} Code will be made available on GitHub upon paper acceptance.
% \footnote{Code will be made available on GitHub upon paper acceptance.} % footnote does not help

\tocless\section{Validation on artificial data}
\label{sec:artificial}

We compare our method with the other commonly used non-parametric algorithms for mutual information estimation: Kraskov-Stögbauer-Grassberger (KSG~\cite{kraskov2004estimating}), Bias-Improved-KSG by Gao et al. (BI-KSG~\cite{gao2016demystifying}) and the Mutual Information Neural Estimator (MINE~\cite{belghazi2018mutual}).

\begin{figure}[ht]
    \centering
    \includegraphics[width=\textwidth]{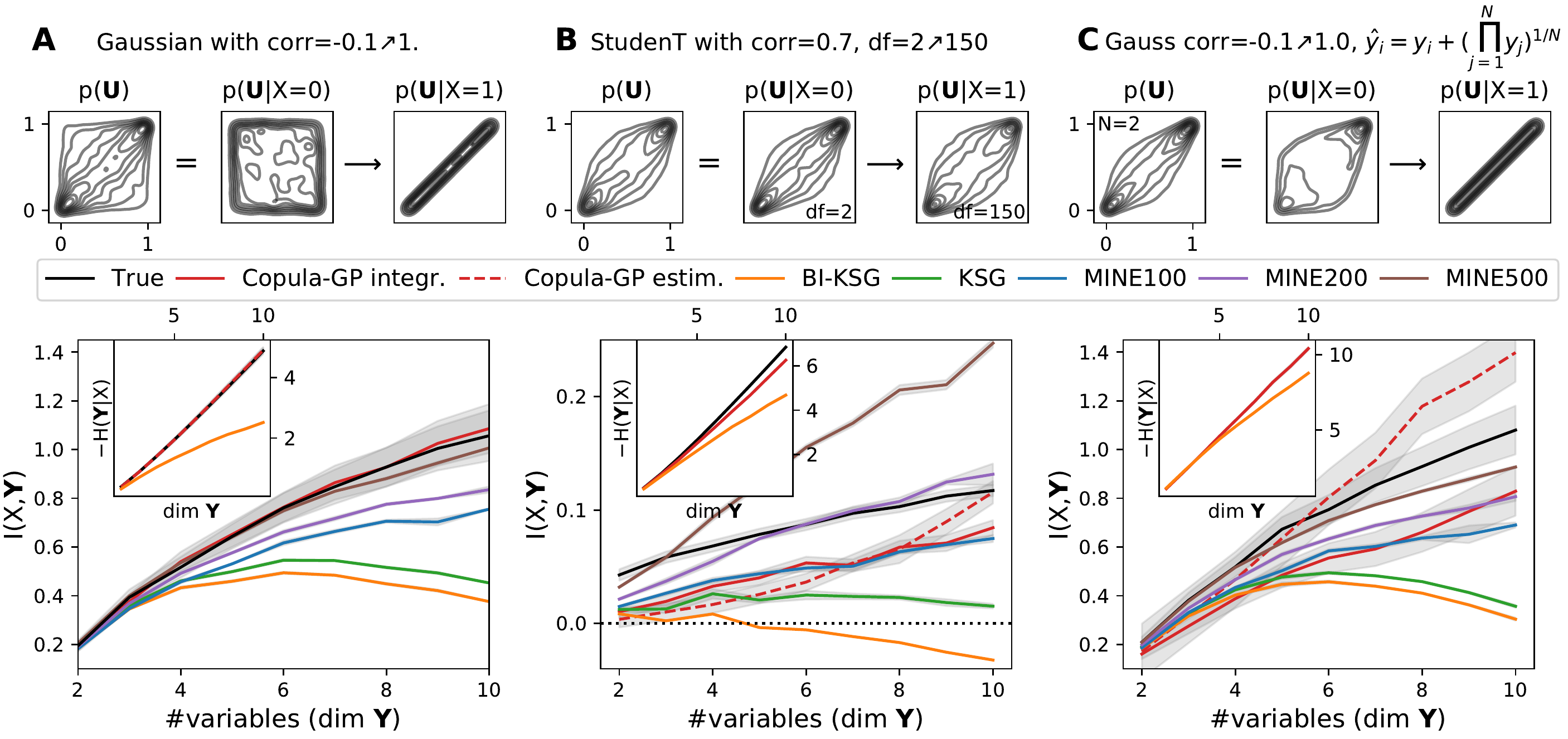}
    \caption{Conditional entropy $H(\bY|X)$ and mutual information $I(X,\bY)$ measured by different methods on synthetic data. 
    Upper row shows the dependency structures $p(\bU)$ %for $\bU =\bF(\bY)$ 
    and conditional dependency structures at the beginning and the end of the $\textrm{dom} X = [0,1]$.
    %Plots in the bottom row: black line -- true value; red -- Copula-GP (solid: MC integration (\ref{eq:MI}); dashed: estimated MI (\ref{eq:MI_alt})); orange -- BI-KSG; green -- KSG; blue -- MINE (100~HU); purple -- MINE (500~HU).
    \textbf{A}\,~Multivariate Gaussian samples.
    %with correlation linearly dependent on~$X$.
    \textbf{B}\,~Multivariate Student~T samples.
    %with constant correlation and degrees of freedom dependent on~$X$ exponentially.
    \textbf{C}\,~Multivariate Gaussian samples $\bY$ (same as \textbf{A}), morphed into another distribution $p(\hat\bY)$ with a tail dependence, while $I(X,\bY) = I(X,\hat\bY)$.
    Gray intervals show either standard error of mean (SE, 5 repetitions), or $\sqrt{(SE)^2  + (MC_{tol})^2}$ for integrated variables.
    }
    \label{fig:synth_res}
\end{figure}

First, we test these estimators on a dataset sampled from a multivariate Gaussian distribution, with $\mathrm{cov}(Y_i,Y_j) = \rho + (1-\rho)\,\delta_{ij}$, where $\delta_{ij}$ is Kronecker’s delta and $\rho = -0.1+1.1 X, X~\in~[0,1]$.
Our algorithm selects a Gaussian copula on these data, which perfectly matches the true distribution. 
Therefore, Copula-GP measures both entropy and mutual information exactly (within integration tolerance, see Fig.~\ref{fig:synth_res}A).
The performance of the non-parametric methods on this dataset is lower.
It was shown before that KSG and MINE both severely underestimate the $\MI$ for high-dimensional Gaussians with high correlation (e.g. see Fig.~1 in \citet{belghazi2018mutual}).
The Copula-GP model (integrated) provides accurate estimates for highly correlated (up to $\rho=0.999$, at least up to 20D) Gaussian distributions (see Supplemental Material).

Next, we test the Copula-GP performance on the Student~T distribution, which can only be approximated by the Copula-GP model, but would not exactly match any of the parametric copula families in Table~\ref{tab:families}.
We keep the correlation coefficient $\rho$ fixed at $0.7$, and only change the number of degrees of freedom from 2 to 150 exponentially: $df = \exp(5X)+1, X \in [0,1]$. 
This makes the dataset particularly challenging, as all of the mutual information $I(X,\bY)$ is encoded in tail dependencies of $p(\bY|X)$.
The true $H(\bY|X)$ of the Student~T distribution was calculated analytically (see Eq.~A.12 in~\cite{calsaverini2009studenth}) and $I(X,\mathbf{Y})$ was integrated numerically according to~(\ref{eq:MI}) given the true $p(\bY|X)$.

Figure~\ref{fig:synth_res}B shows that most of the methods underestimate $I(X,\bY)$. 
Copula-GP (integrated) and MINE (with 100 hidden units) provide the closest estimates. The training curve for MINE with more hidden units (200,500) showed signs of overfitting (abrupt changes in loss at certain permutations) and the resulting estimate was higher than the true $I(X,\bY)$ at higher dimensions.
It was shown before that MINE provides inaccurate and inconsistent results on datasets with low $I(X,\bY)~$\cite{song2019understanding}.
We also demonstrate $I(X,\bY)$ estimation with a combination of two copula models for $H(\bY)$ and $H(\bY|X)$: "Copula-GP estimated" (see Eq.~\ref{eq:MI_alt}). In lower dimensions, it captures less information than "Copula-GP integrated", but starts overestimating the true $\MI$ at higher dimensions, when the inaccuracy of the density estimation for $p(\bY)$ builds up. This shows the limitation of the "estimated" method, which can either underestimate or overestimate the correct value due to parametric model mismatch, whereas "integrated" method consistently underestimates the correct value.
We conclude that Copula-GP and MINE demonstrate similar performance in this example, while KSG-based methods significantly underestimate $I(X,\bY)$ in higher dimensions.

Finally, we created another artificial dataset that is not related to any of the copula models used in our framework (Table~\ref{tab:families}).
We achieved that by applying a homeomorphic transformation $\mathbf{F(Y)}$ to a multivariate Gaussian distribution. 
Since the transformation is independent of the conditioning variable, it does not change the $I(X,\bY) = I(X,\mathbf{F(Y)})$~\cite{kraskov2004estimating}. Therefore, we possess the true $I(X,\bY)$, which is the same as for the first example in Figure~\ref{fig:synth_res}A.
Note, however, that the entropy of the samples changes: $H(\bY) \neq H(\mathbf{F(Y)})$. So, there is no ground truth for the conditional entropy.
% We make the correlation between variables linearly dependent on $X$: $\rho = -0.1 + 1.1 X, X \in [0,1]$.
We transform the Gaussian copula samples 
$\bY\in U_{[0,1]}^N$ from the first example as 
$\widetilde{Y_i} = Y_i + ( \prod_{j=1}^N Y_j )^{1/N}$
and again transform the marginals using the empirical probability integral transform $\bU = \bF(\widetilde\bY$).
Both conditional $p(\bU|X)$ and unconditional $p(\bU)$ densities here do not match any of the parametric copulas from Table~\ref{tab:families}. 
As a result, "Copula-GP estimated" overestimated the correct value, while "Copula-GP integrated" underestimated it similarly to the MINE estimator with 100 hidden units.
%The other methods performed similarly to the Student~T example.
%, because the static copula model has less flexibility than the conditional one and poorly describes the density $p(\bY)$, leading to underestimation of $|H(\bY)|$ and overestimation of $I(X,\bY)$. 

% The last 2 examples show the main limitation of our method. 
% If the dependence structure significantly differs from the set of approximating copula families, and no mixture of approximating copula elements can match the target density, then the method performs poorly.
Figure~\ref{fig:synth_res} demonstrates that the performance of the parametric Copula-GP model critically depends on the match between the true probability density and the best mixture of parametric copula elements. 
When the parametric distribution matches the true distribution~(Fig.~\ref{fig:synth_res}A), our Copula-GP framework predictably outperforms all non-parametric methods.
Nonetheless, even when the exact reconstruction of the density is not possible~(Figs.~\ref{fig:synth_res}B-C),
the mixtures of the copula models~(\ref{eq:mix}) are still able to model the changes in tail dependencies, at least qualitatively.
As a result, our method performs similarly to the neural-network based method (MINE) and still outperforms KSG-like methods. 
%which struggle to estimate information in high dimensional datasets with strongly correlated variables.
% The parametric copulas used in this framework were shown to be effective in describing the dependencies in a variety of realistic datasets~\cite{hernandez2013nips}.

% In the next section, we show that the mixtures of parametric copula families in our framework (Table~\ref{tab:families}) are also suitable for describing neuronal and behavioral dependencies, which in turn promises high performance of Copula-GP on estimating information measures in these data. %EXT?

\tocless\section{Validation on real data}
\label{sec:real}
\begin{figure}[b]
    \centering
    \includegraphics[width=\textwidth]{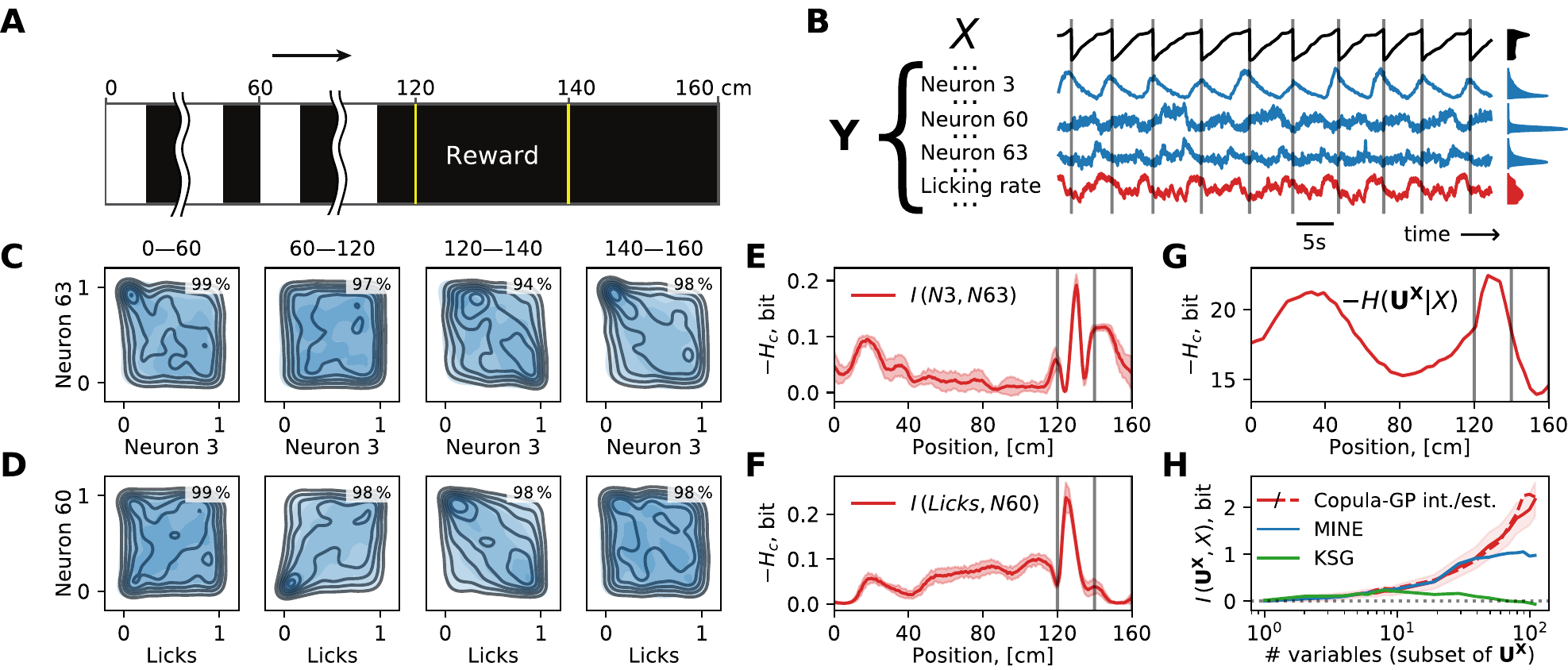}
    \caption{Applications of the Copula-GP framework to neuronal and behavioral data from the visual cortex. 
    \textbf{A}~Schematic of the experimental task~\cite{pakan2018impact,henschke2020reward} in virtual reality (VR);
    \textbf{B}~Example traces from ten example trials: $X$ is a position in VR, $\bY$ is a vector of neuronal recordings (blue) and behavioral variables (red); 
    %Marginal statistics for each variable is shown on the right.
    \textbf{C-D}~Density plots for: 
    the noise correlation (C) and 
    the behavioral modulation (D) examples;
    \textbf{E-G}~Conditional entropy for the
    bivariate examples (E-F) and 
    the population-wide statistics (G);
    \textbf{H}~Comparison of Copula-GP vs. non-parametric estimators on subsets of variables.
    }
    \label{fig:exp_res}
\end{figure}

We investigate the dependencies observed in neuronal and behavioral data and showcase possible applications of the Copula-GP framework.
We used two-photon calcium imaging data of neuronal population activity in the primary visual cortex of mice engaged in a visuospatial navigation task in virtual reality (data from \citet{henschke2020reward}). Briefly, the mice learned to run through a virtual corridor with vertical gratings on the walls (Fig.~\ref{fig:exp_res}A, 0-120cm) until they reached a reward zone (Fig.~\ref{fig:exp_res}A, 120-140cm), where they could get a reward by licking a reward spout.
We condition our Copula-GP model on the position in the virtual environment $X$ and studied the joint distribution of the behavioral ($\tilde{Y}_1\ldots \tilde{Y}_5$) and neuronal ($\tilde{Y}_6 \ldots \tilde{Y}_{109}$) variables ($\mathrm{dim} \bY$=109). 
Figure~\ref{fig:exp_res}B shows a part of the dataset (trials 25-35 out of 130).
The traces here demonstrate changes in the position $X$ of the mouse as well as the activity of 3~selected neurons and the licking rate.
These variables have different patterns of activity depending on $X$ and different signal-to-noise ratios.
Both differences are reflected in marginal statistics, which are shown on the right with the density plots of equal area.

\paragraph{Constructing interpretable bivariate models}
We first studied bivariate relationships between neurons.
In order to do this, we transformed the raw signals (shown in Fig.~\ref{fig:exp_res}B) with a probability integral transform $\bU = \bF(\bY)$.
We observed strong non-trivial changes in the dependence structure $c(\bU|X)$ subject to the position in the virtual reality $X$ and related visual information (Fig.~\ref{fig:exp_res}C). 
% This conditional probability $c(\bU|X)$ describes
% the joint variability of two neuronal signals given a certain stimulus.
Such stimulus-related changes in the joint variability of two neuronal signals are commonly described as \emph{noise correlations}. The Copula-GP model provides a more detailed description of the joint probability that goes beyond linear correlation analysis.
%Using our model, we can calculate the extra information about the visual stimulus: $I(X,\bU) = 0.030\pm0.004$ bit. %EXT? Est better?
In this example, the dependence structure is best characterized by a combination of Gaussian and Clayton copula (rotated by 90$^\circ$).
The density plots Fig.~\ref{fig:exp_res}C demonstrate the match between the true density (outlines) and the copula model density (blue shades) for each part of the task.
We measure the accuracy of the density estimation with the proportion of variance explained $R^2$, which shows how much of the variance of the variable $Y_2$ can be predicted given the variable $Y_1$ (see Eq.(1) in Supplemental Material).
The average $\overline{R^2}$ for all $Y_1$ is provided in the upper right corner of the density plots.

Next, we show that our model can be applied not only to the neuronal data, but also to any of the behavioral variables.
Fig.~\ref{fig:exp_res}D shows the dependence structure between one of the neurons and the licking rate.
%and the mixture of copulas provides interpretable description of the changes in dependence structure.
The best selected mixture model here is Frank + Clayton~0$^\circ$ + Gumbel~270$^\circ$, which again provides an accurate estimate of the conditional dependence between the variables.
Therefore, Figs.~\ref{fig:exp_res}C-D demonstrate that our Copula-GP model provides both an accurate fit for the probability distribution and an interpretable visualization of the dependence structure.

Figs.~\ref{fig:exp_res}E-F show the absolute value of the conditional entropy $|H(\bU^X|X)|$, which is equivalent to the mutual information between two variables $I(U_1^X,U_2^X)$. 
For both examples, the $\MI$ peaks in the reward zone.
The bivariate Copula-GP models were agnostic of the reward mechanism in this task, yet they revealed the position of the reward zone as an anomaly in the mutual information.

\paragraph{Measuring information content in a large neuronal population}

Finally, we constructed a C-vine describing the distribution between all neuronal and behavioral variables ($\mathrm{dim} \bU^X = 109$) and measured the conditional entropy $H(\bU^X|X)$ for all variables in the dataset $\{U^X_1...U^X_{109}\}$.
The conditional entropy in Fig.~\ref{fig:exp_res}G peaks in the reward zone (similarly to Figs.~\ref{fig:exp_res}E-F) and also at the beginning of the trial.
Now the model is informed on the velocity of the animal and the reward events, so the first peak can be attributed to the acceleration of the mouse at the start of a new trial~\cite{khan2018contextual,pakan2018action}.

While constructing the C-vine, we ordered the variables according to their pairwise rank correlations (see Sec.~2.3). %~\ref{sec:vine}
We considered subsets of the first $N$ variables and measured the $\MI$ with the position for each subset.
%We estimate the $I(X,\mathbf{U^X})$ rather than $I(X,\bY)$ here in order to evaluate the performance of our method only, excluding the estimation of marginals. %EXT?
We compared the performance of our Copula-GP method on these subsets of $\mathbf{U^X}$ vs. KSG and MINE.
Fig.~\ref{fig:exp_res}H shows that all 3 methods provide similar results on subsets of up to 10~variables, yet in higher dimensions both MINE and KSG show smaller $I(X,\{U_{i<N}^X\})$ compared to our Copula-GP method, which agrees with the results obtained on the synthetic data (Fig.~\ref{fig:synth_res}).
The true values of $I(X,\{U_{i<N}^X\})$ are unknown, yet we expect the integrated Copula-GP (solid line) to underestimate the true value due to parametric model mismatch.
The Copula-GP "estimated" (dashed line) almost perfectly matches the "integrated" result, which suggests that the model was able to accurately approximate both $p(\bU^X|X)$ and $p(\bU^X)$, and, as a result, $I(X,\{U_{i<N}^X\})$.
These results demonstrate superior performance of our Copula-GP model on high-dimensional neuronal data.
% The last variables in the ordered dataset are the neurons, that are weakly correlated with the rest of the population. Yet, ...
% We have seen in the synthetic data examples (Fig.~\ref{fig:synth_res}), both MINE and KSG often underestimate the value of $\MI(X,\bY)$ in high dimensions.
% Unlike the synthetic data, neuronal recordings have sparse and weak dependencies, yet we observe similar behavior on this dataset as well.

\tocless\section{Discussion}
    We have developed a Copula-GP framework for modeling conditional multivariate joint distributions.
    The method is based on linear mixtures of parametric copulas, which provide flexibility for estimating complex dependencies.
    We approximate conditional dependencies of the model parameters with Gaussian Processes which allow us to implement a Bayesian model selection procedure.
    The selected models combine the accuracy in density estimation with the interpretability of parametric copula models.
    %with qualitatively different properties (e.g. tail dependencies).
    Despite the limitations of the parametric models, our framework demonstrated good results in mutual information estimation on the synthetically generated data, performing similarly to the state-of-the-art non-parametric information estimators.
    The framework is also well suited for describing neuronal and behavioral data.
    The possible applications include, but are not limited to, studying noise correlations, behavioral modulation and the neuronal population statistics.
    % In terms of scalability, the bottleneck of our method is the MC integration that we used for information estimates.
    We demonstrated that the model scales well at least up to 109 variables, while theoretically, the parameter inference scales as $\mathcal{O}(n\cdot m^2)$, where $n$ is a number of samples and $m$ is the (effective) number of variables (see Suppl. Mat.).
    %, since scaling of the MC to higher dimensions may become infeasible. %, even though the sparsity and weak connectivity of the neuronal data facilitated the MC convergence.
    % We also proposed an approximate method for estimating mutual information, which scales...
    In summary, we demonstrated that the Copula-GP approach can make stochastic relationships explicit
    and generate accurate and interpretable models of dependencies between neuronal responses, sensory stimuli, and behavior. 
    Future work will focus on implementing model selection for the vine structure and improving the scalability of the $\MI$ estimation algorithm.

    % Scales well. Hierarchical, modular algorithm, easy to parallelise.

    % Cyclic domain, multi-dimensional conditioning variable.
    
    % In terms of scalability, the bottleneck of our method is MC integration 

\section*{Broader Impact}

We envision a wide range of impacts resulting from the use of the Copula-GP framework in computational neuroscience as well as in machine learning research, information technology and economics.

In computational neuroscience, this approach has the potential to reveal high dimensional context-dependent relationships between neuronal activity and behavioral variables. Therefore, our model can provide novel insights into the principles of neural circuit-level computation, both under physiological conditions, and during the aberrant network function observed in many neuropsychiatric disorders~\cite{uhlhaas2012neuronal,bassett2018understanding,braun2018maps}. Furthermore, understanding context-dependent processing in the brain might suggest ways to mimic the same principles in artificial neural networks. 

The proposed framework also has some potentially far reaching applications, not only in computational neuroscience but also in information technology and economics. Current problems of information flow in computer networks as well as high-speed trading in micro- and macro-markets can be phrased as non-stationary relationship problems that require proper stochastic representations, which, in turn, can benefit economic success. On the other hand, there is a possible risk of a one-sided adoption of the method by malicious actors benefiting from market manipulation, which may give them unfair advantage and thus reduce the market transparency and cause economic damage.

We encourage the researchers adapting our method to understand the limitations of the use of parametric copula models. The choice of the bivariate copula models and the vine structures introduces our beliefs about the particular conditional dependency or independency into the model. The misuse of parametric copula models has once had a negative impact on the insurance industry in the past~\cite{donnelly_embrechts_2010}. 
Thus, these limitations must be taken into consideration when designing new applications in the future. 

One potential negative societal impact related to the use of our framework may include a relatively large energy footprint from training the models on graphics processing units (GPUs). The models used for creating figures in this paper took about 2 weeks of computational time on 8 GPUs. The mitigation strategy should be focused on careful planning of the simulations, creating checkpoints and backups to prevent data loss, reuse of the trained models and other practices that reduce the energy-consuming computation.

% Authors are required to include a statement of the broader impact of their work, including its ethical aspects and future societal consequences. 
% Authors should discuss both positive and negative outcomes, if any. For instance, authors should discuss a) 
% who may benefit from this research, b) who may be put at disadvantage from this research, c) what are the consequences of failure of the system, and d) whether the task/method leverages
% biases in the data. If authors believe this is not applicable to them, authors can simply state this.

% Use unnumbered first level headings for this section, which should go at the end of the paper. {\bf Note that this section does not count towards the eight pages of content that are allowed.}

\begin{ack}
We thank the GENIE Program and the Janelia Research Campus, specifically
V. Jayaraman, R. Kerr, D. Kim, L. Looger, and K. Svoboda, for making GCaMP6 available. This
work was funded by the Engineering and Physical Sciences Research Council (grant
EP/S005692/1 to A.O.), the Wellcome Trust and the Royal Society (Sir Henry Dale fellowship to N.R.),
the Marie Curie Actions of the European Union’s FP7 program (MC-CIG 631770 to N.R.), the
Simons Initiative for the Developing Brain (to N.R.), the Precision Medicine Doctoral Training
Programme (MRC, the University of Edinburgh (to T.A.)).
%and by the Wellcome Trust and the Royal Society (Sir Henry Dale fellowship to N.R.)

\end{ack}
% \discretionary{-}{-}{-}
% \section*{References}
\small
\bibliographystyle{unsrtnat} % temporarily use this style
\bibliography{references}   % do not want to care about ref order while writing

\clearpage
% \begin{center}
%     \bf\Large
%     Supplemental Material for ``Parametric Copula-GP model for analyzing multidimensional neuronal and behavioral relationships''
% \end{center}
\title{Supplemental Material for ``Parametric Copula-GP model for analyzing multidimensional neuronal and behavioral relationships''}
\setcounter{section}{0}
\setcounter{equation}{0}
\setcounter{table}{0}
\renewcommand*{\thesection}{S\arabic{section}}

\makeatletter
\renewcommand{\thefigure}{S\@arabic\c@figure}
\renewcommand{\theequation}{S\@arabic\c@equation}
\renewcommand{\thetable}{S\@arabic\c@table}
\makeatother

\def\ALGSTYLE#1{\mathtt{#1}}

\def\bY{\mathbf{Y}}
\def\bU{\mathbf{U}}
\def\WAIC{\mathrm{WAIC}}
\def\lppd{\mathrm{lppd}}
\def\GPU{RTX 2080Ti}
\def\MI{\mathrm{MI}}
\def\Mw{M_{\text{worst}}}
\def\Mb{M_{\text{best}}}

\maketitle

\tableofcontents

\vspace{0.7cm}
\noindent\rule{\textwidth}{0.4pt}
\vspace{0.1cm}

\section{Methods}

\subsection{Goodness-of-fit}

We measure the accuracy of the density estimation with the proportion of variance explained $R^2$.
We compare the empirical conditional CDF $\mathrm{ecdf}(U_2|U_1=y)$ vs. estimated conditional CDF $\mathrm{ccdf}(U_2|U_1=y)$ and calculate:
\begin{equation}
    R^2(y) = 1 - \sum_{U_2} \left(\frac{\mathrm{ecdf}(U_2|U_1=y) - \mathrm{ccdf}(U_2|U_1=y)}{\mathrm{ecdf}(U_2|U_1=y) - \overline{U_2}}\right)^2,
    \label{eq:S1}
\end{equation}
where $R^2(y)$ quantifies the portion of the total variance of $U_2$ that our copula model can explain given $U_1=y$, and $\overline{U_2} = \overline{F(Y_2)} = 0.5$. The sum was calculated for $U_2 = 0.05 n, n=0\ldots20$.

Next, we select all of the samples from a certain interval of the task ($X \in [X_1,X_2]$) matching one of those shown in Figure~3 in the paper. 
We split these samples $U_1 \in [0,1]$ into 20 equally sized bins: $\{I_i\}_{20}$.
For each bin $I_i$, we calculate (\ref{eq:S1}).
We evaluate $\mathrm{ccdf}(U_2|U_1=y_i) \approx \mathrm{ccdf}(U_2|U_1\in I_i)$ using a copula model from the center of mass of the considered interval of $X$: $X_{\mu} = \mathrm{mean}(X)$ for samples $X \in [X_1,X_2]$.
We use the average measure: 
\begin{equation}
    \overline{R^2} = \mathop{\mathbb{E}}_{p(U_1\in I_i)} R^2\big(\mathrm{mean}(U_1\in I_i)\big),
    \label{eq:S2}
\end{equation}
to characterize the goodness of fit for a bivariate copula model.
Since $U_1$ is uniformly distributed on $[0,1]$, the probabilities for each bin $p(U_1\in I_i)$ are equal to $1/20$, and the resulting measure $\overline{R^2}$ is just an average $R^2$ from all bins.
The results were largely insensitive to the number of bins (e.g. 20 vs. 100).

\subsection{Variational inference}

Since our model has latent variables with GP priors and intractable posterior distribution, the direct maximum likelihood Type-II estimation is not possible and an approximate inference is needed.
We used stochastic variational inference (SVI) with a single evidence lower bound~\citeS{hensman2015scalable}:
\begin{equation}
    \mathcal{L}_\mathrm{ELBO} =
    \sum_{i=1}^N \mathop{\mathbb{E}}_{q(f_i)} \big[\log p(y_i|f_i)\big]
    - \mathop{\mathrm{KL}}[q(\mathbf{u})||p(\mathbf{u})],
    \label{eq:ELBO}
\end{equation}
implemented as \texttt{VariationalELBO} in GPyTorch~\citeS{gardner2018gpytorchS}.
Here $N$ is the number of data samples, $\mathbf{u}$ are the inducing points, $q(\mathbf{u})$ is the variational distribution and $q(\mathbf{f}) = \int p(\mathbf{f}|\mathbf{u})q(\mathbf{u})d\mathbf{u}$. 

Following the \citetS{wilson2015kernelS} approach (KISS-GP), we then constrain the inducing points to a regular grid, which applies a deterministic relationship between $\mathbf{f}$ and $\mathbf{u}$. 
As a result, we only need to infer the variational distribution $q(\mathbf{u})$, but not the positions of $\mathbf{u}$. The number of grid points is one of the model hyper-parameters: \texttt{grid\_size}.

Equation~\ref{eq:ELBO} enables joint optimization of the GP hyper-parameters (constant mean $\mu$ and two kernel parameters: scale and bandwidth) and parameters of the variational distribution $\mathbf{q}$ (mean and covariance at the inducing points: $\mathbf{u} \sim \mathcal{N} (\mu_u \times \mathbf{1},\Sigma_u)$ )~\citeS{hensman2015scalable}.
We have empirically discovered by studying the convergence on synthetic data, that the best results are achieved when the learning rate for the GP hyper-parameters (\texttt{base\_lr}) is much greater than the learning rate for the variational distribution parameters (\texttt{var\_lr}, see Table~\ref{tab:hypp}).

\paragraph{Priors} For both the neuronal and the synthetic data, we use a standard normal prior $p(\mathbf{u}) \sim \mathcal{N} (\mathbf{0},I)$ for a variational distribution.
Note, that the parametrization for mixture models was chosen such that the aforementioned choice of the variational distribution prior with zero mean corresponds to \emph{a priori} equal mixing coefficients $\phi_j = 1/M$ for $j=1\ldots M$.
In our experiments with the simulated and real neuronal data, we observed that the GP hyper-parameter optimisation problem often had 2 minima (which is a common situation, see Figure 5.5 on page 116 in \citetS{williams2006gaussian}). %, corresponding to different $\lambda$, $\sigma_n^2$. 
One of those corresponds to a short kernel lengthscale ($\lambda$) and low noise ($\min_\mathbf{f} \sigma^2$), which we interpret as overfitting.
To prevent overfitting, we used $\lambda \sim \mathcal{N} (0.5,0.2)$ prior on RBF kernel lengthscale parameter %, using the fact that all the inputs in all experiments were normalized and constrained on $X \in [0,1]$. 
that allows the optimizer to approach the minima from the region of higher $\lambda$, ending up in the minimum with a larger lengthscale.

\paragraph{Optimization} We use the Adam optimizer with two learning rates for GP hyper-parameters (\texttt{base\_lr}) and variational distribution parameters (\texttt{var\_lr}).
We monitor the loss (averaged over 50 steps) and its changes in the last 50 steps: \texttt{$\Delta$ loss = mean(loss[-100:-50]) - mean(loss[-50:])}. 
If the change becomes smaller than \texttt{check\_waic}, then we evaluate the model $\WAIC$ and check if it is lower than $-\WAIC_{tol}$. 
If it is higher, we consider that either the variables are independent, or the model does not match the data.
Either way, this indicates that further optimisation is counterproductive.
If the $\WAIC<-\WAIC_{tol}$, we proceed with the optimisation until the change of loss in 50 steps $\Delta loss$ becomes smaller than \texttt{loss\_tol} (see Table~\ref{tab:hypp}).

\paragraph{Effective learning rates for different families}

The coefficients in the GPLink functions for different copula families are also a part of model hyper-parameters. 
The choice of these coefficients affects the gradients of the log probability function. 
Since GPLink functions are nonlinear, they affect the gradients in various parameter ranges to a different extent.
This results in variable convergence rates depending on the true copula parameters.

To address the problem of setting up these hyper-parameters, we have created the tests on synthetic data with different copula parameters.
Using these tests, we manually adjusted these hyper-parameters such that the GP parameter inference converged in around 1000-2000 iterations for every copula family and parameter range.
We have also multiplied the GP values corresponding to the mixture coefficients by $0.5$, to effectively slow down the learning of the mixture coefficients $\phi$ compared to the copula coefficients $\theta$, which also facilitates the convergence.

\paragraph{Hyper-parameter selection} 
The hyper-parameters of our model (Table~\ref{tab:hypp}) were manually tuned, often considering the trade off between model accuracy and evaluation time.
A more systematic hyper-parameter search might yield improved results and better determine the limits of model accuracy.

\begin{table}[h]
  \caption{Hyper-parameters of the bivariate Copula-GP model}
  \label{tab:hypp}
  \centering
  \begin{tabular}{lll}
    \toprule
    % & \multicolumn{2}{c}{Domain}                   \\
    % \cmidrule(r){2-3}
    Hyper-parameter  &    Value      & Description \\
    \midrule
    \texttt{base\_lr}         &   $10^{-2}$   & Learning rate for GP parameters \\
    \texttt{var\_lr}          &   $10^{-3}$   & Learning rate for variational distribution \\
    \texttt{grid\_size}       &   128         & Number of inducing points for KISS-GP \\
    \texttt{waic\_tol}        &   0.005       & Tolerance for $\WAIC$ estimation \\
    \texttt{loss\_tol}        &   $10^{-4}$   & Loss tolerance that indicates the convergence \\
    \texttt{check\_waic}      &   0.005       & Loss tolerance when we check $\WAIC$ \\
    \multicolumn{3}{l}{$\ldots$ and GPLink parameters listed in Table~1.}\\
    \bottomrule
  \end{tabular}
\end{table}

\subsection{Bayesian model selection}

In model selection, we are aiming to construct a model with the lowest possible $\WAIC$.
Since our copula probability densities are continuous, their values can exceed 1 and the resulting $\WAIC$ is typically negative. Zero $\WAIC$ corresponds to the Independence model ($\mathrm{pdf} = 1$ on the whole unit square). 
We also set up a tolerance ($\WAIC_{tol} = 0.005$), and models with $\WAIC \in [-\WAIC_{tol},\WAIC_{tol}]$ are considered indistinguishable from the independence model.

Since the total number of combinations of 10~copula elements (Fig.1) is large, exhaustive search for the optimal model is not feasible.
In our framework, we propose two model algorithms for constructing close-to-optimal copula mixtures:
\emph{greedy} and \emph{heuristic}.

\subsection{Model selection algorithms}

\begin{algorithm}
    \def\Size{\ALGSTYLE{size}}
    \def\Reduce{\ALGSTYLE{reduce}}
    \def\Prepend{\ALGSTYLE{prepend}}
    
    \def\MinimalModelSize{\ALGSTYLE{minimal\_model\_size}}
    \def\Independence{\mathtt{Independence}}
    \def\Gauss{\ALGSTYLE{Gauss}}
    \def\Frank{\ALGSTYLE{Frank}}
    \def\Clayton{\ALGSTYLE{Clayton}}
    \def\Gumbel{\ALGSTYLE{Gumbel}}
    \caption{Greedy algorithm for copula mixture selection}
    \label{alg:greedy}
    \def\Mo{M_{\text{old}}}
    \def\Mb{M_{\text{best}}}
    \def\Sc{S_{\text{c}}}
    $M, \Mo \leftarrow [~], [~]$\;
    $\Sc \leftarrow [\Independence, \Gauss, \Frank, 4\times\Clayton, 4\times\Gumbel]$\;
    
    \tcp*[f]{4$\times$ includes all rotations}
    
    \tcp{while every update of the model yields a new best}
    \While{$\WAIC(M) \leq \WAIC(\Mo)$ {\rm\bf and} $\Size(\Sc)>0$}
    {
        $\Mo \leftarrow M$ \;
        select  $c$ from $\Sc$ such that $\WAIC(\Prepend(c,M))$ is minimal\;
        $M\leftarrow \Prepend(c,M)$ \;
        remove $c$ from $\Sc$\;
        %  \textbf{if} $\WAIC(M) < \WAIC(\Mb)$\;
        % \If{$\WAIC(M) < \WAIC(\Mb)$}{
        % $\Mb \leftarrow M$
        % }
    }
    $\Mb \leftarrow \Reduce(\Mo)$\;
    
    \KwRet$\Mb$\;
\end{algorithm}

\begin{algorithm}
    \def\Prepend{\ALGSTYLE{prepend}}
    \def\Reduce{\ALGSTYLE{reduce}}
    \def\Size{\ALGSTYLE{size}}
    \SetKw{Break}{break}
    
    \def\WAICtol{\mathtt{waic\_tol}}
    \def\Independence{\mathtt{Independence}}
    \def\Gauss{\ALGSTYLE{Gauss}}
    \def\Frank{\ALGSTYLE{Frank}}
    \def\Clayton{\ALGSTYLE{Clayton}}
    \def\Gumbel{\ALGSTYLE{Gumbel}}
    
    \caption{Heuristic algorithm for copula mixture selection}
    \label{alg:heuristic}
    $G\leftarrow [\Gauss]$\;
    \If{$\WAIC(G)>-\WAICtol$}{
        \KwRet $[\Independence]$\;
    }
    $M_{Cl} \leftarrow [\Independence,\Gauss,4 \times \Clayton]$\;
    $M_{Gu} \leftarrow [\Independence,\Gauss,4 \times \Gumbel]$\;
    % important \leftarrow [$M_{Cl}.mix[i]>10\%$ {\rm\bf or} $M_{Gu}.mix[i]>10\%$~{\rm\bf for}~i~{\rm\bf in}~1\ldots 4] \;
    $\Mb, \Mw \leftarrow$ ($M_{Cl}$,$M_{Gu}$) {sorted by} $\WAIC$\;
    
    \If{$\WAIC(G)<\WAIC(\Mb)$}{
        \KwRet $G$\;
    }

    \For{$i\leftarrow 3\ldots\Size(\Mb)$}{
        $M\leftarrow\Mb$ with $i$-th element replaced by $\Mw[i]$\;
        $\Mb \leftarrow M$ \textbf{if} $\WAIC(M) < \WAIC(\Mb)$\;
        % \If{$\WAIC(M) < \WAIC(\Mb)$}{
        %     $\Mb \leftarrow M$\;
        % }
    }
    
    $\Mb \leftarrow \Reduce(\Mb)$\;
    \If{$\Gauss\in\Mb$}{
        $M\leftarrow\Mb$ with $\Gauss$ replaced by $\Frank$\;
        $\Mb \leftarrow M$ \textbf{if} $\WAIC(M) < \WAIC(\Mb)$\;
    }
    \tcp{$\Gauss$ often gets confused with pairs of e.g. $\Clayton\,0^\circ + \Gumbel\,0^\circ$}
    
    \If{$\Size(\Mb)>1$}{
        \For{$i\leftarrow 1\ldots(\Size(\Mb)-1)$}{
            \For{$j\leftarrow (i+1)\ldots\Size(\Mb)$}{
                $M\leftarrow\Mb$ with $i$-th and $j$-th elements removed\;
                $M\leftarrow \Prepend(\Gauss,M)$\;
                \If{$\WAIC(M)<\WAIC(\Mb)$}{
                    $\Mb \leftarrow M$ \;
                    \Break\;
                }
            }
        }
    }
    
    $\Mb \leftarrow \Reduce(\Mb)$\;
    
    \KwRet $\Mb$\;

\end{algorithm}

The greedy algorithm (Algorithm~\ref{alg:greedy}) starts by comparing $\WAIC$ of all possible single-copula models (from Table~1, in all rotations) and selecting the model with the lowest $\WAIC$. After that, we add one more copula (from another family or in another rotation) to the first selected copula, and prepend the element that yields the lowest $\WAIC$ of the mixture. We repeat the process 
%at least one more time (3-element mixtures) and then
until the $\WAIC$ stops decreasing.
After the best model is selected, we remove the inessential elements using the \texttt{reduce(.)} function. This function removes those elements which have an average concentration of $<10\%$ everywhere on $X\in[0,1]$. This step is added to improve the interpretability of the models and computation time for entropy estimation (at a small accuracy cost) and can, in principle, be omitted.
% The selection process takes approximately (30-60)~minutes$\times$number of copulas in the correct mixture on a~\GPU~, but terminates in less than 1~minute for independent variables. 

The greedy algorithm can be improved by adding model reduction after each attempt to add an element. In this case, the number of elements can increase and decrease multiple times during the model selection process, which also must be terminated if the algorithm returns to the previously observed solution. Even though it complicates the algorithm, it reduces the maximal execution time (observed on the real neuronal data) from $\sim$90~minutes down to $\sim$40~minutes.
% Another way to speed up model selection is adding a condition after evaluating the Gaussian copula.

The heuristic algorithm focuses on the tail dependencies (Algorithm~\ref{alg:heuristic}).
First, we try a single Gaussian copula.
%, which is the one that is often selected first by the greedy algorithm as well. 
If variables are not independent, we next compare 2 combinations of 6 elements, which are organized as follows: 
an Independence copula together with a Gaussian copula and either 4~Clayton or 4~Gumbel copulas in all 4~rotations (0$^\circ$, 90$^\circ$, 180$^\circ$, 270$^\circ$). 
We select the combination with the lowest $\WAIC$. After that, we take the remaining Clayton/Gumbel copulas one by one and attempt to switch the copula type (Clayton to Gumbel or vise versa). If this switching decreases the $\WAIC$, we keep a better copula type for that rotation and proceed to the next element.

Here we make the assumption, that because Clayton and Gumbel copulas have most of the probability density concentrated in one corner of the unit square (the heavy tail), we can choose the best model for each of the 4 corners independently.
When the best combination of Clayton/Gumbel copulas is selected, we can (optionally) reduce the model.

We have not yet used a Frank copula in a heuristic algorithm.
We attempt to substitute the Gaussian copula with a Frank copula (if it is still a part of the reduced mixture, see lines 16-19 in Alg.~\ref{alg:heuristic}).
%, or a combination of Clayton \& Gumbel or two Gumbel copulas. 
Sometimes, a Gaussian copula can be mistakenly modeled as a Clayton \& Gumbel or two Gumbel copulas.
So, as a final step (lines 20-31, Alg.~\ref{alg:heuristic}), we select all pairwise combinations of the remaining elements, and attempt to substitute each of the pairs with a Gaussian copula, selecting the model with the lowest $\WAIC$.
Despite a large number of steps in this algorithm, the selection process takes only up to 25 minutes (in case all elements in all rotations are required).

The procedure was designed after observing the model selection process on a variety of synthetic and real neuronal datasets. 
%It can be improved by conducting a systematic study of the confusion matrix for the copula mixtures. One can generate samples from combinations of a few (1-3) copulas and check the $\WAIC$ for all combinations of 1-3 element models. Yet, this would require considerable computational time.

\subsection{Vine copulas}
\label{sec:vines}

Vine models provide a way to factorize the high-dimensional copula probability density into a hierarchical set of bivariate copulas~\citeS{aas2009pairS}.
There are many possible decompositions based on different assumptions about conditional independence of specific elements in a model, which can be classified using graphical models called \emph{regular vines}~\citeS{bedford2001probability,bedford2002vines}. 
A regular vine can be represented using a hierarchical set of trees, where each node corresponds to a conditional distribution function (e.g. $F(U_2|U_1)$) and each edge corresponds
to a bivariate copula (e.g. $c(U_2,U_3|U_1)$).
The copula models from the lower trees are used to obtain new conditional distributions (new nodes) with additional conditional dependencies for the higher trees, e.g. a \texttt{ccdf} of a copula $c(U_2,U_3|U_1)$ and a marginal conditional distribution $F(U_2|U_1)$ from the 1st tree provide a new conditional distribution $F(U_3|U_1,U_2)$ for a 2nd tree. 
Therefore, bivariate copula parameters are estimated sequentially, starting from the lowest tree and moving up the hierarchy.
The total number of edges in all trees (= the number of bivariate copula models) for an $m$-dimensional regular vine equals $m(m-1)/2$.

The regular vines often assume that the conditional copulas $c(U_i,U_j|\{U_k\})$ themselves are independent of their conditioning variables $\{U_k\}$, but depend on the them indirectly through the conditional distribution functions (nodes)~\citeS{acar2012beyond}. 
This is known as the \emph{simplifying assumption} for vine copulas~\citeS{haff2010simplified}, which, if applicable, allows to escape the curse of dimensionality in high-dimensional copula construction.

In this study, we focus on the \emph{canonical vine} or \emph{C-vine}, which has a unique node in each tree, connected to all of the edges in that tree. For illustration, see, for example, Figure~2 in~\citetS{aas2009pair}.
The C-vine was shown to be a good choice for neuronal datasets~\citeS{onken2016nipsS}, as they often include some proxy of neuronal population activity as an outstanding variable, strongly correlated with the rest.
This variable provides a natural choice for the first conditioning variable in the lowest tree.
In the neuronal datasets from~\citetS{henschke2020rewardS}, this outstanding variable is the global fluorescence signal in the imaged field of view (global neuropil).

To construct a C-vine for describing the neuronal and behavioural data from~\citetS{henschke2020rewardS}, we used a heuristic element ordering based on the sum of absolute values of Kendall's $\tau$ of a given element with all of the other elements.
It was shown by~\citetS{czado2012maximumS} that this ordering facilitates C-vine modeling.
For all of the animals and most of the recordings (14 out of 16), including the one used in Figure~3, the first variable after such ordering was the global neuropil activity.
This again confirms, that a C-vine with the global neuropil activity as a first variable is an appropriate model for the dependencies in neuronal datasets.

% Justify the use of C-Vine, explain that the first variable is always the neuropil activity for~\citet{pakan2018impact} data, no matter what day/animal, cite~\cite{onken2016nips}.

\subsection{Algorithmic complexity}
\label{sec:algcomp}

In this section, we discuss the algorithmic complexity of the parameter inference for a C-vine copula model.

The parameter inference for each of the bivariate Copula-GP models scales as $\mathcal{O}(n)$, where $n$ is the number of samples, since we use a scalable kernel interpolation KISS-GP~\citeS{wilson2015kernel}.
As we mentioned in Sec.~\ref{sec:vines}, a full $m$-dimensional C-vine model requires $m(m-1)/2$ bivariate copulas, trained sequentially.
As a result, the $\mathcal{O}(n)$ GP parameter inference has to be repeated $m(m-1)/2$ times, which yields $\mathcal{O}(n\cdot m^2)$ complexity.

In practice, the computational cost (in terms of time) of the parameter inference for each bivariate model varies from tens of seconds to tens of minutes.
The heuristic model selection is designed in such a way, that it discards independent variables in just around 20 seconds  (line~3 in Alg.~\ref{alg:heuristic}).
As a result, most of the models are quickly skipped and further considered as Independence models, and their contribution to the total computational cost can be neglected.
When the model is evaluated, the Independence components are also efficiently `skipped' during sampling, as \texttt{ppcf} function is not called for them.
The Independence models also add zero to C-vine log probability, so they are also `skipped' during log probability calculation.
They also reduce the total memory storage, as no GP parameters, which predominate the memory requirements, are stored for these models.

In a conditional C-vine trained on a real neuronal dataset with 109 variables, 5253 out of 5886 (89\%) bivariate models were Independence, which leaves
only 633 non-Independence models.

In practice, this means that the algorithmic complexity of the model is much better than the na\"ive theoretical prediction $\mathcal{O}(n\cdot m^2)$, based on the structure of the graphical model. 
Suppose that the actual number of the non-Independence models $N_{nI}$ in a vine model is much smaller than $m(m-1)/2$ and can be characterized by an effective number of dimensions $m_{eff} \sim \sqrt{N_{nI}}$. 
In this case, instead of the $\mathcal{O}(m^2)$ scaling with the number of variables, the complexity highly depends on the sparsity of the dependencies in the graphical model and scales with as $\mathcal{O}(n\cdot N_{nI}) \sim \mathcal{O}(n\cdot m_{eff}^2)$.

Therefore, the our method is especially efficient on the datasets with a low effective dimensionality $m_{eff}$, such as the neuronal data. The number of variables $m$ itself has little effect on the computational cost and memory storage.
% This can be shown as follows. 
% Suppose we transformed the dataset $Y$ with the PCA or a similar technique minimising the information loss. This yields a new dataset $\widetilde{Y}$ with the most of the information (variance) in the first $m_{eff}$ components.

\section{More validation on synthetic data}

\paragraph{Computing infrastructure} We developed our framework and ran the majority of our experiments (described both in the paper and Supplemental Material) on an Ubuntu 18.04 LTS machine with 2 x Intel(R) Xeon(R) Gold 6142 CPU @ 2.60GHz and 1x GeForce~RTX~2080 + 1 x GeForce~RTX~2080~Ti GPUs.
For training C-vine models, we used another Scientific Linux 7.6 machine with 1 x Intel(R) Xeon(R) Silver 4114 CPU @ 2.20GHz and 8 x GeForce~RTX~2080~Ti GPUs.

\paragraph{Code availability} Code will be made available on GitHub upon paper acceptance.

\subsection{Model selection for bivariate copulas}

\paragraph{Synthetic data}
We generate artificial data by sampling from a copula mixture, parametrized in two different ways: 
\begin{enumerate}
\item mixing concentrations of all copulas were constant and equal to $1/N$ ($N$ = number of copulas), but copula parameters $\theta$ were parametrized by the phase-shifted sinus functions:
\begin{equation}
\label{eq:sin}
\theta_i = A_i\sin\left(\pi m\frac{i}{N} + 2\pi x\right) + B_i, \qquad x \in [0,1]
\end{equation}
where $i$ is the index of the copula in a mixture, $m=1$. For Clayton and Gumbel copulas, the absolute value of the sinus was used. 
The amplitudes $A_i$ were chosen to cover most of the range of parameters, except for extremely low or high $\theta$s for which all copula families become indistinguishable (from independence or deterministic dependence, respectively). 
\item copula parameters $\theta$ were constant, but mixing concentrations $\phi$ were parametrized by the phase-shifted sinus functions (same as Eq.~\ref{eq:sin}, with $A_i = B_i = 1/N$ and $m=2$). 
Such parametrization ensures that the sum of all mixing concentrations remains equal to one ($\sum_{i=1}^N\phi = 1$). 
Yet, each $\phi$ turns to zero somewhere along this trajectory, allowing us to discriminate the models and infer the correct mixture. 
\end{enumerate}

\paragraph{Identifiability tests}
We tested the ability of the model selection algorithms to select the correct mixture of copula models, the same as the one from which the data was generated.
We generated 5000 samples with equally spaced unique inputs on [0,1]. %how to describe linspace(0,1,5000) better?

Both model selection algorithms were able to correctly select all of the 1-component and most of the 2-component models on simulated data. 
For simulated data with larger numbers of components (or 2 very similar components), the $\WAIC$ of the selected model was either lower (which is possible given a limited number of samples) or close to the $\WAIC$ of the correct parametric model. 
In other words, the difference between the $\WAIC$ of the correct model and of the best selected model never exceeded the $\WAIC_{test\_tol}=0.05$, which we set up as a criteria for passing the test: $\Delta\WAIC<\WAIC_{test\_tol}$.
Since all the tests were passed successfully, we conclude that both algorithms are capable of finding optimal or close-to-optimal solutions for copula mixtures.

\paragraph{A more detailed report on the model identifiability tests}

Tables~\ref{tab:search1}-\ref{tab:search3T} below illustrate the search for the best model.
The copula model names in these tables are shortened to the first two letters, e.g. Gumbel becomes `Gu', Frank becomes `Fr'.
The information in these Tables provides some intuition on the model selection process and the range of $\WAIC$s for the correct or incorrect models. The final selected models are shown in bold.

Table~\ref{tab:search1} demonstrates that both greedy and heuristic algorithms can identify the correct single copula model.
Some key intermediate models ($M$ in Alg.~\ref{alg:greedy}-\ref{alg:heuristic}) with their $\WAIC$s are listed in the table, along with the total duration of simulations (T, in minutes) on~\GPU~ for both algorithms.

Table~\ref{tab:search2M} shows the identification of the mixtures with 2 components, where the copula parameters $\theta$ were constant (independent of $X$) and mixing concentrations $\phi$ were parameterized by the phase-shifted sinus functions (Eq.~\ref{eq:sin}).
All of these models were correctly identified with both algorithms.
The mixtures with 2 components, where the copula parameters $\theta$ varied harmonically (as in Eq.~\ref{eq:sin}) but the mixing concentrations $\phi$ were constant, were harder to identify.
Table~\ref{tab:search2T} shows that a few times, each of the algorithms selected a model that was better than the true model ($\WAIC_{best}-\WAIC_{true}<0$).
The greedy algorithm made one mistake, yet the model it selected was very close to optimal. Such misidentification happens due to the limited number of samples in a given synthetic dataset.

Tables~\ref{tab:search3M}-\ref{tab:search3T} show the model selection for 3 component models. Again, as in Tables~\ref{tab:search2M}-\ref{tab:search2T}, either $\theta$ or $\phi$ was constant. Here, the model selection algorithms could rarely identify the correct model (due to overcompleteness of the mixture models), but always selected the one that was very close to optimal: $\WAIC_{best}-\WAIC_{true} \ll \WAIC_{test\_tol}$.

Note, that $\WAIC_{test\_tol}$ is different from \texttt{waic\_tol}.
We have set \texttt{waic\_tol} for comparison against Independent model to such a small value (10x smaller than $\WAIC_{test\_tol}$) because we want to avoid making false assumptions about conditional independences in the model.
Also note, that the $\WAIC$ of the true model depends on the particular synthetic dataset generated in each test. Therefore, the final $\WAIC$ in the left and in the right columns of Tables~\ref{tab:search1}-\ref{tab:search3T} can be slightly different (yet, right within $\WAIC_{test\_tol}$).

% \subsection{Model selection for C-vine}

% Using the same bivariate copula parametrization, we generated a 5-variable model. A C-vine tree for 5-variables contains 10 bivariate conditional copulas. We have randomly assigned a unique copula type to each of the 

\subsection{Accuracy of entropy estimation}

\begin{figure}[htb]
    \centering
    \includegraphics[width=\textwidth]{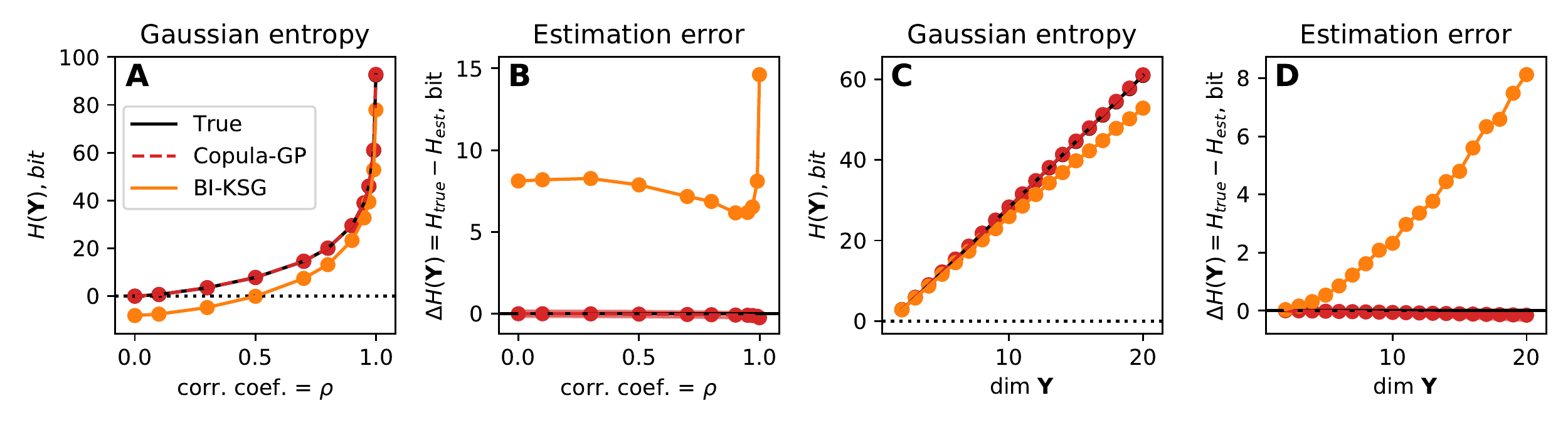}
    \caption{Accuracy of the entropy estimation for multivariate Gaussian distributions.
    \textbf{A}~Entropy of the 20-dimensional multivariate Gaussian distribution for different correlation coefficients $\rho$.
    \textbf{B}~Estimation error for the entropy shown in A.
    \textbf{C}~Entropy of the multivariate Gaussian distributions with $\rho=0.99$ and varying dimensionality.
    \textbf{D}~Estimation error for the entropy shown in C.
    }
    \label{fig:gauss}
\end{figure}

In this section, we consider a fixed copula mixture model with known parameters $\theta$ and test the reliability of the entropy estimation with Monte Carlo (MC) integration.
We test the accuracy of the entropy estimation on a multivariate Gaussian distribution, with $\mathrm{cov}(Y_i,Y_j) = \rho + (1-\rho)\,\delta_{ij}$, where $\delta_{ij}$ is Kronecker’s delta and $\rho \in~[0,0.999]$.
Given a known Gaussian copula, we estimate the entropy with MC integration and compare it to the analytically calculated true value.
We set up a tolerance to $\Delta H = 0.01(\mathrm{dim} \bY)$.
As a result, for every correlation $\rho$ (Fig.~\ref{fig:gauss}A-B) and every number of dimensions $\mathrm{dim} \bY$ (Fig.~\ref{fig:gauss}C-D), the Copula-GP provides an accurate result, within the error margin.
In Figure~\ref{fig:gauss}, BI-KSG estimates~\citeS{gao2016demystifyingS} obtained on the dataset with 10k samples are shown for comparison.
This experiment 1) validates the MC integration;
2) validates the numerical stability of the probability density function of the Gaussian copula up to a specified maximal $\rho=0.999$ (for $\rho>0.999$ the model is indistinguishable from the deterministic dependence $U_1=U_2$).

% \section{Results on neuronal data}

\section{Model parameters for the bivariate neuronal and behavioural examples}

\begin{figure}[htb]
    \centering
    \includegraphics[width=\textwidth]{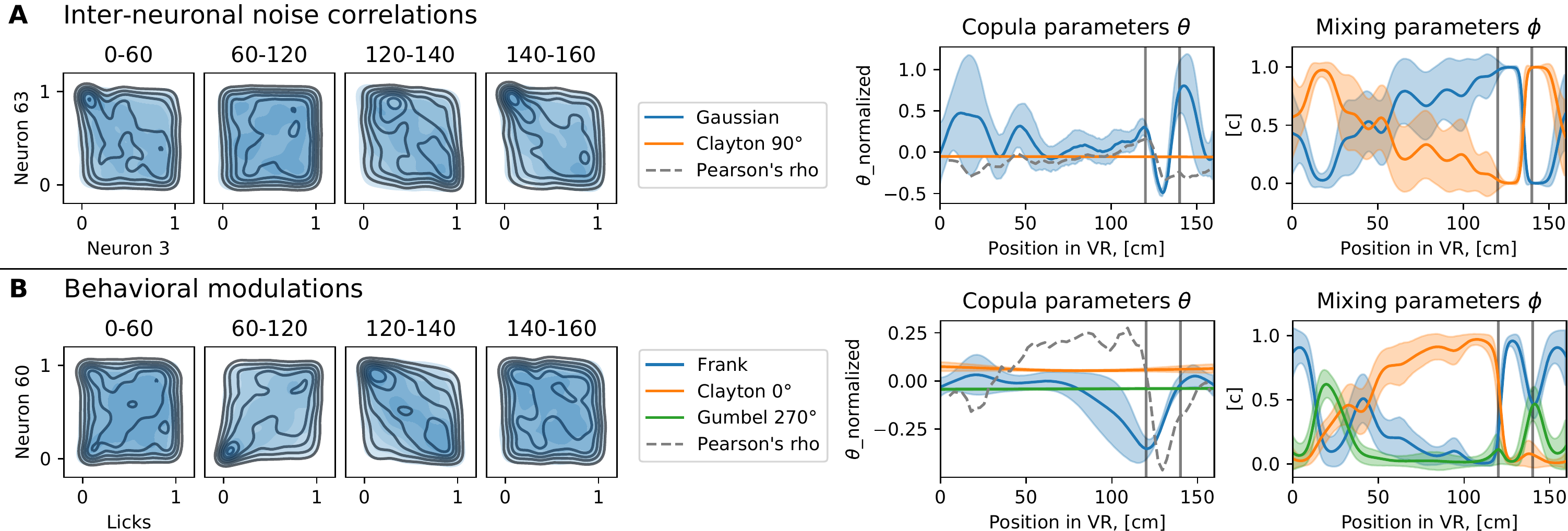}
    \caption{Parameters of the copula mixture models.
    From left to right: copula probability densities (same as Fig.3C-D); a list of selected copula elements; copula parameters $\theta$; mixing concentrations $\phi$.
    These plots are provided for:
    \textbf{A}~the noise correlation example;
    \textbf{B}~the behavioral modulation example. 
    }
    \label{fig:tuning}
\end{figure}

In this section, we provide visualisations for the parameters of the bivariate copula models from Figure~3C-F and discuss the interpretability of these models.

Figure~\ref{fig:tuning} shows the probability density of the joint distribution of two variables and the parameters of a corresponding Copula-GP mixture model.
The plots on the left repeat Fig.3C-D and represent the true density (outlines) and the copula model density (blue shades) for each part of the task.

In the noise correlation example (Fig.~\ref{fig:tuning}A), we observe the \emph{tail dependencies} between the variables (i.e. concentration of the probability density in a corner of the unit square) around [0-60]~cm and [140-160]~cm of the virtual corridor.
There is only one element with a tail dependency in this mixture: Clayton~90$^\circ$ copula.
On the right-most plot in Fig.~\ref{fig:tuning}A, we see the mixing concentration for the elements of the mixture model.
The concentration of Clayton~90$^\circ$ copula (orange line) is close to 100\% around 20~cm and 150~cm, which agrees with our observations from the density plots.

The confidence intervals ($\pm2\sigma$) for the parameters approximated with Gaussian processes are shown with shaded areas in parameter plots.
These intervals provide a measure of uncertainty in model parameters.
For instance, when the concentration of the Gaussian copula in the mixture is close to 0\% ($X$ around 20~cm and 150~cm), the confidence intervals for the Gaussian copula parameter ($\theta$, blue shade) in Fig.~\ref{fig:tuning}A become very wide (from almost 0 to 1).
Since this copula element is not affecting the mixture for those values of $X$, its $\theta$ parameter has no effect on the mixture model log probability.
Therefore, this parameter is not constrained to any certain value.
In a similar manner, we see that the variables are almost independent between 60 and 120~cm (see density plots on the left in Fig.~\ref{fig:tuning}). Both copula elements can describe this independence.
As a result, the mixing concentrations for both elements have high uncertainty in that interval of $X$.
Yet, Gaussian copula with a slightly positive correlation is still a bit more likely to describe the data in that interval.

The copula parameter plot in Fig.~\ref{fig:tuning}A also shows Pearson's $\rho$, which does not change much in this example and remains close to zero. This illustrates, that the traditional linear noise correlation analysis would ignore (or downplay) this pair of neurons as the ones with no dependence.
This happens because the Pearson's $\rho$ only captures the linear correlation and ignores the tail dependencies, whereas our model provides a more detailed description of the joint bivariate distribution.

In the behavioural modulation example (Fig.~\ref{fig:tuning}B), we observe more complicated tail dependencies in the density plots.
The best selected model supports this observation and provides a mixture model with 3 components, 2 of which have various tail dependencies.
The Clayton~0$^\circ$ copula (orange) describes the lower tail dependence observed in the second part of the virtual corridor with gratings (around [60-120]~cm, see Fig.~3A for task structure).
This dependence can be verbally interpreted as follows:
\emph{when there is \textbf{no} licking, the Neuron~60 is certainly silent; but when the animal \textbf{is} licking, the activity of Neuron~60 is slightly positively correlated with the licking rate}.

These examples illustrate, that by analysing the copula parameters and the mixing concentrations of the Copula-GP mixture model, one can interpret the changes in the bivariate dependence structure.
Just like traditional \emph{tuning curves} characterize the response of a single neuron, our mixture model characterizes the `tuning' of the dependence structure between pairs of variables to a given stimulus or context.
Knowing the qualitative properties of the copula elements that constitute a copula mixture, one can focus on the dominant element of the copula mixture for every given conditioning variable $X$ and describe the shape of the dependence.

\small
\bibliographystyleS{unsrtnat} % temporarily use this style
\bibliographyS{references_Suppl_1}   % do not want to care about ref order while writing

% \begin{table}[h]
\def\maketabletitle{True & \multicolumn{3}{c|}{Greedy} & \multicolumn{3}{c}{Heuristic}                   \\
    Model &    Search attempts      & WAIC & T &  Search attempts      & WAIC & T \\}

  \newpage
  \small
  \captionof{table}{The model selection histories for 1-element mixtures}
  \label{tab:search1}
  \addtocounter{table}{-1}
  \centering
  \begin{longtable}{l|p{3.3cm}ll|p{3.45cm}ll}
    \toprule
    \maketabletitle
    \midrule
    \endhead
    Ga              & Ga                             & -0.1619 & 25\,m & Ga                             & -0.1513 & 3\,m\\*
                & Ga\-Fr                         & -0.1610 &       & In\-Ga\-Gu$^{180}$\-Gu$^{270}$\-Gu$^{0}$\-Gu$^{90}$ & -0.1499 &      \\*
                & \textbf{Ga}                    & \textbf{-0.1619} &       & In\-Ga\-Cl$^{0}$\-Cl$^{90}$\-Cl$^{180}$\-Cl$^{270}$ & -0.1498 &      \\*
                &                                &  &       & \textbf{Ga}                    & \textbf{-0.1513} &      \\

\midrule
Fr              & Fr                             & -0.1389 & 57\,m & Ga                             & -0.1400 & 3\,m\\*
                & Fr\-Cl$^{90}$                  & -0.1395 &       & In\-Ga\-Gu$^{180}$\-Gu$^{270}$\-Gu$^{0}$\-Gu$^{90}$ & -0.1391 &      \\*
                & Fr\-Cl$^{90}$\-Gu$^{270}$      & -0.1396 &       & In\-Ga\-Cl$^{0}$\-Cl$^{90}$\-Cl$^{180}$\-Cl$^{270}$ & -0.1391 &      \\*
                & Fr\-Cl$^{90}$\-Gu$^{270}$\-Gu$^{90}$ & -0.1396 &       & \textbf{Fr}                    & \textbf{-0.1509} &      \\*
                & \textbf{Fr}                    & \textbf{-0.1389} &       &                                &  &      \\

\midrule
Cl$^{0}$        & Cl$^{0}$                       & -0.5225 & 37\,m & Ga                             & -0.3825 & 5\,m\\*
                & Cl$^{0}$\-Gu$^{0}$             & -0.5226 &       & In\-Ga\-Gu$^{180}$\-Gu$^{270}$\-Gu$^{0}$\-Gu$^{90}$ & -0.4943 &      \\*
                & Cl$^{0}$\-Gu$^{0}$\-Cl$^{180}$ & -0.5225 &       & In\-Ga\-Cl$^{0}$\-Cl$^{90}$\-Cl$^{180}$\-Cl$^{270}$ & -0.5303 &      \\*
                & \textbf{Cl$^{0}$}              & \textbf{-0.5224} &       & \textbf{Cl$^{0}$}              & \textbf{-0.5311} &      \\

\midrule
Gu$^{0}$        & Gu$^{0}$                       & -0.6267 & 43\,m & Ga                             & -0.5555 & 7\,m\\*
                & Gu$^{0}$\-Cl$^{180}$           & -0.6268 &       & In\-Ga\-Gu$^{180}$\-Gu$^{270}$\-Gu$^{0}$\-Gu$^{90}$ & -0.5988 &      \\*
                & Gu$^{0}$\-Cl$^{180}$\-Gu$^{180}$ & -0.6267 &       & In\-Ga\-Cl$^{0}$\-Cl$^{90}$\-Cl$^{180}$\-Cl$^{270}$ & -0.5946 &      \\*
                & \textbf{Gu$^{0}$}              & \textbf{-0.6230} &       & Ga\-Gu$^{0}$                   & -0.6040 &      \\*
                &                                &  &       & \textbf{Gu$^{0}$}              & \textbf{-0.6050} &      \\

\midrule
Cl$^{90}$       & Cl$^{90}$                      & -0.5389 & 22\,m & Ga                             & -0.3922 & 5\,m\\*
                & Cl$^{90}$\-Cl$^{270}$          & -0.5389 &       & In\-Ga\-Gu$^{180}$\-Gu$^{270}$\-Gu$^{0}$\-Gu$^{90}$ & -0.5047 &      \\*
                & \textbf{Cl$^{90}$}             & \textbf{-0.5389} &       & In\-Ga\-Cl$^{0}$\-Cl$^{90}$\-Cl$^{180}$\-Cl$^{270}$ & -0.5409 &      \\*
                &                                &  &       & \textbf{Cl$^{90}$}             & \textbf{-0.5410} &      \\

\midrule
Gu$^{90}$       & Gu$^{90}$                      & -0.6137 & 55\,m & Ga                             & -0.5501 & 7\,m\\*
                & Gu$^{90}$\-Gu$^{270}$          & -0.6144 &       & In\-Ga\-Gu$^{180}$\-Gu$^{270}$\-Gu$^{0}$\-Gu$^{90}$ & -0.5893 &      \\*
                & Gu$^{90}$\-Gu$^{270}$\-Cl$^{270}$ & -0.6145 &       & In\-Ga\-Cl$^{0}$\-Cl$^{90}$\-Cl$^{180}$\-Cl$^{270}$ & -0.5831 &      \\*
                & Gu$^{90}$\-Gu$^{270}$\-Cl$^{270}$\-Cl$^{90}$ & -0.6144 &       & Ga\-Gu$^{90}$                  & -0.5887 &      \\*
                & \textbf{Gu$^{90}$}             & \textbf{-0.6137} &       & \textbf{Gu$^{90}$}             & \textbf{-0.5950} &      \\

\midrule
Cl$^{180}$      & Cl$^{180}$                     & -0.5566 & 36\,m & Ga                             & -0.3932 & 7\,m\\*
                & Cl$^{180}$\-Cl$^{0}$           & -0.5582 &       & In\-Ga\-Gu$^{180}$\-Gu$^{270}$\-Gu$^{0}$\-Gu$^{90}$ & -0.4956 &      \\*
                & Cl$^{180}$\-Cl$^{0}$\-In       & -0.5582 &       & In\-Ga\-Cl$^{0}$\-Cl$^{90}$\-Cl$^{180}$\-Cl$^{270}$ & -0.5493 &      \\*
                & \textbf{Cl$^{180}$}            & \textbf{-0.5565} &       & \textbf{Cl$^{180}$}            & \textbf{-0.5489} &      \\

\midrule
Gu$^{180}$      & Gu$^{180}$                     & -0.6131 & 43\,m & Ga                             & -0.5553 & 6\,m\\*
                & Gu$^{180}$\-Cl$^{0}$           & -0.6164 &       & In\-Ga\-Gu$^{180}$\-Gu$^{270}$\-Gu$^{0}$\-Gu$^{90}$ & -0.6091 &      \\*
                & Gu$^{180}$\-Cl$^{0}$\-Fr       & -0.6163 &       & In\-Ga\-Cl$^{0}$\-Cl$^{90}$\-Cl$^{180}$\-Cl$^{270}$ & -0.6045 &      \\*
                & \textbf{Gu$^{180}$}            & \textbf{-0.6131} &       & \textbf{Gu$^{180}$}            & \textbf{-0.6154} &      \\

\midrule
Cl$^{270}$      & Cl$^{270}$                     & -0.5434 & 23\,m & Ga                             & -0.3909 & 5\,m\\*
                & Cl$^{270}$\-Gu$^{270}$         & -0.5433 &       & In\-Ga\-Gu$^{180}$\-Gu$^{270}$\-Gu$^{0}$\-Gu$^{90}$ & -0.5094 &      \\*
                & \textbf{Cl$^{270}$}            & \textbf{-0.5434} &       & In\-Ga\-Cl$^{0}$\-Cl$^{90}$\-Cl$^{180}$\-Cl$^{270}$ & -0.5535 &      \\*
                &                                &  &       & \textbf{Cl$^{270}$}            & \textbf{-0.5548} &      \\

\midrule
Gu$^{270}$      & Gu$^{270}$                     & -0.5928 & 51\,m & Ga                             & -0.5763 & 6\,m\\*
                & Gu$^{270}$\-Cl$^{90}$          & -0.5934 &       & In\-Ga\-Gu$^{180}$\-Gu$^{270}$\-Gu$^{0}$\-Gu$^{90}$ & -0.6277 &      \\*
                & Gu$^{270}$\-Cl$^{90}$\-In      & -0.5935 &       & In\-Ga\-Cl$^{0}$\-Cl$^{90}$\-Cl$^{180}$\-Cl$^{270}$ & -0.6179 &      \\*
                & Gu$^{270}$\-Cl$^{90}$\-In\-Cl$^{180}$ & -0.5931 &       & \textbf{Gu$^{270}$}            & \textbf{-0.6300} &      \\*
                & \textbf{Gu$^{270}$}            & \textbf{-0.5928} &       &                                &  &      \\

    \bottomrule
  \end{longtable}
  
  \newpage
  \captionof{table}{The model selection histories for 2-element mixtures with constant $\theta$ and variable $\phi$}
  \label{tab:search2M}
  \addtocounter{table}{-1}
  \centering
  \begin{longtable}{p{0.8cm}|p{3.3cm}ll|p{3.45cm}ll}
    \toprule
    \maketabletitle
    \midrule
    \endhead
    Gu$^{90}$      & Ga                             & -0.1877 & 101\,m & Ga                             & -0.1922 & 11\,m\\*
Ga            & Ga\-Gu$^{90}$                  & -0.2855 &       & In\-Ga\-Gu$^{180}$\-Gu$^{270}$\-Gu$^{0}$\-Gu$^{90}$ & -0.3070 &      \\*
                & Ga\-Gu$^{90}$\-Cl$^{270}$      & -0.2855 &       & In\-Ga\-Cl$^{0}$\-Cl$^{90}$\-Cl$^{180}$\-Cl$^{270}$ & -0.2996 &      \\*
                & Ga\-Gu$^{90}$\-Cl$^{270}$\-Fr  & -0.2856 &       & Ga\-Cl$^{0}$\-Gu$^{0}$\-Gu$^{90}$ & -0.3082 &      \\*
                & Ga\-Gu$^{90}$\-Cl$^{270}$\-Fr\-Gu$^{270}$ & -0.2856 &       & Ga\-Cl$^{0}$\-Cl$^{180}$\-Gu$^{90}$ & -0.3076 &      \\*
                & Ga\-Gu$^{90}$\-Cl$^{270}$\-Fr\-Gu$^{270}$\-Cl$^{90}$ & -0.2856 &       & \textbf{Ga\-Gu$^{90}$}         & \textbf{-0.3091} &      \\*
                & \textbf{Gu$^{90}$\-Ga}         & \textbf{-0.2854} &       &                                &  &      \\

\midrule
Ga            & Fr                             & -0.1635 & 87\,m & Ga                             & -0.1600 & 5\,m\\*
Cl$^{270}$    & Fr\-Cl$^{270}$                 & -0.2707 &       & In\-Ga\-Gu$^{180}$\-Gu$^{270}$\-Gu$^{0}$\-Gu$^{90}$ & -0.2687 &      \\*
                & Fr\-Cl$^{270}$\-Ga             & -0.2747 &       & In\-Ga\-Cl$^{0}$\-Cl$^{90}$\-Cl$^{180}$\-Cl$^{270}$ & -0.2835 &      \\*
                & Fr\-Cl$^{270}$\-Ga\-Gu$^{180}$ & -0.2782 &       & \textbf{Ga\-Cl$^{270}$}        & \textbf{-0.2845} &      \\*
                & Fr\-Cl$^{270}$\-Ga\-Gu$^{180}$\-Cl$^{90}$ & -0.2781 &       &                                &  &      \\*
                & \textbf{Ga\-Cl$^{270}$}        & \textbf{-0.2821} &       &                                &  &      \\

\midrule
Gu$^{180}$    & Gu$^{180}$                     & -0.1681 & 99\,m & Ga                             & -0.1534 & 8\,m\\*
Fr              & Gu$^{180}$\-Fr                 & -0.2099 &       & In\-Ga\-Gu$^{180}$\-Gu$^{270}$\-Gu$^{0}$\-Gu$^{90}$ & -0.1993 &      \\*
                & Gu$^{180}$\-Fr\-Cl$^{180}$     & -0.2101 &       & In\-Ga\-Cl$^{0}$\-Cl$^{90}$\-Cl$^{180}$\-Cl$^{270}$ & -0.1977 &      \\*
                & Gu$^{180}$\-Fr\-Cl$^{180}$\-Cl$^{90}$ & -0.2105 &       & In\-Ga\-Gu$^{180}$             & -0.2074 &      \\*
                & Gu$^{180}$\-Fr\-Cl$^{180}$\-Cl$^{90}$\-In & -0.2106 &       & \textbf{Fr\-Gu$^{180}$}        & \textbf{-0.2104} &      \\*
                & Gu$^{180}$\-Fr\-Cl$^{180}$\-Cl$^{90}$\-In\-Gu$^{270}$ & -0.2099 &       &                                &  &      \\*
                & \textbf{Fr\-Gu$^{180}$}        & \textbf{-0.2099} &       &                                &  &      \\

\midrule
Cl$^{0}$         & Fr                             & -0.1587 & 92\,m & Ga                             & -0.1652 & 5\,m\\*
Cl$^{90}$        & Fr\-Cl$^{0}$                   & -0.2600 &       & In\-Ga\-Gu$^{180}$\-Gu$^{270}$\-Gu$^{0}$\-Gu$^{90}$ & -0.3142 &      \\*
                & Fr\-Cl$^{0}$\-Cl$^{90}$        & -0.3173 &       & In\-Ga\-Cl$^{0}$\-Cl$^{90}$\-Cl$^{180}$\-Cl$^{270}$ & -0.3430 &      \\*
                & Fr\-Cl$^{0}$\-Cl$^{90}$\-Gu$^{270}$ & -0.3176 &       & \textbf{Cl$^{0}$\-Cl$^{90}$}   & \textbf{-0.3448} &      \\*
                & Fr\-Cl$^{0}$\-Cl$^{90}$\-Gu$^{270}$\-In & -0.3176 &       &                                &  &      \\*
                & Fr\-Cl$^{0}$\-Cl$^{90}$\-Gu$^{270}$\-In\-Cl$^{270}$ & -0.3175 &       &                                &  &      \\*
                & \textbf{Cl$^{90}$\-Cl$^{0}$}   & \textbf{-0.3190} &       &                                &  &      \\

\midrule
Cl$^{180}$       & Fr                             & -0.2204 & 103\,m & Ga                             & -0.1965 & 7\,m\\*
Gu$^{270}$        & Fr\-Cl$^{180}$                 & -0.3488 &       & In\-Ga\-Gu$^{180}$\-Gu$^{270}$\-Gu$^{0}$\-Gu$^{90}$ & -0.3591 &      \\*
                & Fr\-Cl$^{180}$\-Gu$^{270}$     & -0.3874 &       & In\-Ga\-Cl$^{0}$\-Cl$^{90}$\-Cl$^{180}$\-Cl$^{270}$ & -0.3688 &      \\*
                & Fr\-Cl$^{180}$\-Gu$^{270}$\-Cl$^{90}$ & -0.3877 &       & Ga\-Gu$^{270}$\-Cl$^{180}$ & -0.3771 &      \\*
                & Fr\-Cl$^{180}$\-Gu$^{270}$\-Cl$^{90}$\-Ga & -0.3878 &       & \textbf{Gu$^{270}$\-Cl$^{180}$} & \textbf{-0.3772} &      \\*
                & Fr\-Cl$^{180}$\-Gu$^{270}$\-Cl$^{90}$\-Ga\-Gu$^{90}$ & -0.3878 &       &                                &  &      \\*
                & \textbf{Gu$^{270}$\-Cl$^{180}$} & \textbf{-0.3888} &       &                                &  &      \\

    \bottomrule
  \end{longtable}
  
  \captionof{table}{The model selection histories for 2-element mixtures with constant $\phi$ and variable $\theta$}
  \label{tab:search2T}
  \addtocounter{table}{-1}
  \centering
  \begin{longtable}{p{0.8cm}|p{3.3cm}ll|p{3.45cm}ll}
    \toprule
    \maketabletitle
    \midrule
    \endhead
    Gu$^{90}$      & Gu$^{90}$                      & -0.1419 & 60\,m & Ga                             & -0.1538 & 10\,m\\*
Ga            & Gu$^{90}$\-Fr                  & -0.2022 &       & In\-Ga\-Gu$^{180}$\-Gu$^{270}$\-Gu$^{0}$\-Gu$^{90}$ & -0.2320 &      \\*
                & Gu$^{90}$\-Fr\-Cl$^{270}$      & -0.2024 &       & In\-Ga\-Cl$^{0}$\-Cl$^{90}$\-Cl$^{180}$\-Cl$^{270}$ & -0.2218 &      \\*
                & Gu$^{90}$\-Fr\-Cl$^{270}$\-Ga  & -0.2024 &       & Ga\-Cl$^{90}$\-Gu$^{0}$\-Gu$^{90}$ & -0.2321 &      \\*
                & \textbf{Fr\-Gu$^{90}$}         & \textbf{-0.2021} &       & \textbf{Ga\-Gu$^{90}$}         & \textbf{-0.2326} &      \\*
                & \hfill $\WAIC_{best} - \WAIC_{true}$:        & \textbf{-0.0013} &       &                                &  &      \\

\midrule
Ga            & Gu$^{90}$                      & -0.1495 & 56\,m & Ga                             & -0.1062 & 7\,m\\*
Cl$^{270}$    & Gu$^{90}$\-Fr                  & -0.1894 &       & In\-Ga\-Gu$^{180}$\-Gu$^{270}$\-Gu$^{0}$\-Gu$^{90}$ & -0.1747 &      \\*
                & Gu$^{90}$\-Fr\-Cl$^{270}$      & -0.1915 &       & In\-Ga\-Cl$^{0}$\-Cl$^{90}$\-Cl$^{180}$\-Cl$^{270}$ & -0.1783 &      \\*
                & Gu$^{90}$\-Fr\-Cl$^{270}$\-In  & -0.1902 &       & Ga\-Gu$^{0}$\-Cl$^{270}$ & -0.1812 &      \\*
                & \textbf{Cl$^{270}$\-Fr\-Gu$^{90}$} & \textbf{-0.1915} &       & \textbf{Ga\-Cl$^{270}$}        & \textbf{-0.1801} &      \\*
                & \hfill $\WAIC_{best} - \WAIC_{true}$:        & \textbf{0.0032} &       &                                &  &      \\

\midrule
Gu$^{180}$    & Gu$^{180}$                     & -0.1600 & 58\,m & Ga                             & -0.1331 & 8\,m\\*
Fr             & Gu$^{180}$\-Fr                 & -0.2191 &       & In\-Ga\-Gu$^{180}$\-Gu$^{270}$\-Gu$^{0}$\-Gu$^{90}$ & -0.1944 &      \\*
                & Gu$^{180}$\-Fr\-Cl$^{270}$     & -0.2195 &       & In\-Ga\-Cl$^{0}$\-Cl$^{90}$\-Cl$^{180}$\-Cl$^{270}$ & -0.1936 &      \\*
                & Gu$^{180}$\-Fr\-Cl$^{270}$\-Cl$^{0}$ & -0.2190 &       & Ga\-Gu$^{180}$\-Cl$^{90}$\-Gu$^{0}$\-Gu$^{90}$ & -0.1945 &      \\*
                & \textbf{Fr\-Gu$^{180}$}        & \textbf{-0.2190} &       & \textbf{Ga\-Gu$^{180}$}        & \textbf{-0.1992} &      \\*
                &                                &  &       & \hfill $\WAIC_{best} - \WAIC_{true}$:        & \textbf{-0.0094} &      \\

\midrule
Cl$^{0}$         & Gu$^{180}$                     & -0.0253 & 62\,m & Ga                             & -0.0079 & 5\,m\\*
Cl$^{90}$        & Gu$^{180}$\-Cl$^{90}$          & -0.2383 &       & In\-Ga\-Gu$^{180}$\-Gu$^{270}$\-Gu$^{0}$\-Gu$^{90}$ & -0.1904 &      \\*
                & Gu$^{180}$\-Cl$^{90}$\-Cl$^{0}$ & -0.2506 &       & In\-Ga\-Cl$^{0}$\-Cl$^{90}$\-Cl$^{180}$\-Cl$^{270}$ & -0.2330 &      \\*
                & Gu$^{180}$\-Cl$^{90}$\-Cl$^{0}$\-In & -0.2509 &       & \textbf{Cl$^{0}$\-Cl$^{90}$}   & \textbf{-0.2361} &      \\*
                & Gu$^{180}$\-Cl$^{90}$\-Cl$^{0}$\-In\-Fr & -0.2508 &       &                                &  &      \\*
                & \textbf{Cl$^{0}$\-Cl$^{90}$}   & \textbf{-0.2586} &       &                                &  &      \\
\midrule
Cl$^{180}$   & Gu$^{270}$                     & -0.0242 & 69\,m & Ga                             & -0.0083 & 6\,m\\*
Gu$^{270}$    & Gu$^{270}$\-Cl$^{180}$         & -0.2499 &       & In\-Ga\-Gu$^{180}$\-Gu$^{270}$\-Gu$^{0}$\-Gu$^{90}$ & -0.2277 &      \\*
                & Gu$^{270}$\-Cl$^{180}$\-Gu$^{180}$ & -0.2517 &       & In\-Ga\-Cl$^{0}$\-Cl$^{90}$\-Cl$^{180}$\-Cl$^{270}$ & -0.2535 &      \\*
                & Gu$^{270}$\-Cl$^{180}$\-Gu$^{180}$\-In & -0.2518 &       & \textbf{Ga\-Cl$^{90}$\-Cl$^{180}$} & \textbf{-0.2549} &      \\*
                & Gu$^{270}$\-Cl$^{180}$\-Gu$^{180}$\-In\-Cl$^{0}$ & -0.2518 &       &                                &  &      \\*
                & Gu$^{270}$\-Cl$^{180}$\-Gu$^{180}$\-In\-Cl$^{0}$\-Fr & -0.2518 &       &                                &  &      \\*
                & \textbf{Cl$^{180}$\-Gu$^{270}$} & \textbf{-0.2500} &       &                                &  &      \\*
                &                                &  &       & \hfill $\WAIC_{best} - \WAIC_{true}$:        & \textbf{-0.0098} &      \\

    \bottomrule
  \end{longtable}
  
  \captionof{table}{The model selection histories for 3-element mixtures with constant $\theta$ and variable $\phi$}
  \label{tab:search3M}
  \addtocounter{table}{-1}
  \centering
  \begin{longtable}{p{0.8cm}|p{3.3cm}ll|p{3.45cm}ll}
    \toprule
    \maketabletitle
    \midrule
    \endhead
    Ga           & Gu$^{0}$                       & -0.1399 & 44\,m & Ga                             & -0.1252 & 6\,m\\*
Cl$^{90}$    & Gu$^{0}$\-Cl$^{90}$            & -0.2494 &       & In\-Ga\-Gu$^{180}$\-Gu$^{270}$\-Gu$^{0}$\-Gu$^{90}$ & -0.2481 &      \\*
Gu$^{0}$        & Gu$^{0}$\-Cl$^{90}$\-Cl$^{0}$  & -0.2519 &       & In\-Ga\-Cl$^{0}$\-Cl$^{90}$\-Cl$^{180}$\-Cl$^{270}$ & -0.2565 &      \\*
                & Gu$^{0}$\-Cl$^{90}$\-Cl$^{0}$\-Fr & -0.2518 &       & \textbf{Ga\-Cl$^{90}$\-Cl$^{180}$} & \textbf{-0.2564} &      \\*
                & \textbf{Cl$^{90}$\-Gu$^{0}$}   & \textbf{-0.2494} &       &                                &  &      \\*
                & \hfill $\WAIC_{best} - \WAIC_{true}$:        & \textbf{-0.0036} &       & \hfill $\WAIC_{best} - \WAIC_{true}$:        & \textbf{0.0014} &      \\

\midrule
Fr               & Fr                             & -0.0591 & 77\,m & Ga                             & -0.0489 & 6\,m\\*
Cl$^{90}$        & Fr\-Cl$^{90}$                  & -0.1460 &       & In\-Ga\-Gu$^{180}$\-Gu$^{270}$\-Gu$^{0}$\-Gu$^{90}$ & -0.1573 &      \\*
Gu$^{0}$        & Fr\-Cl$^{90}$\-Gu$^{0}$        & -0.1730 &       & In\-Ga\-Cl$^{0}$\-Cl$^{90}$\-Cl$^{180}$\-Cl$^{270}$ & -0.1578 &      \\*
                & Fr\-Cl$^{90}$\-Gu$^{0}$\-Cl$^{180}$ & -0.1736 &       & \textbf{Ga\-Cl$^{90}$\-Cl$^{180}$} & \textbf{-0.1621} &      \\*
                & Fr\-Cl$^{90}$\-Gu$^{0}$\-Cl$^{180}$\-In & -0.1734 &       &                                &  &      \\*
                & \textbf{Gu$^{0}$\-Cl$^{90}$\-Fr} & \textbf{-0.1731} &       &                                &  &      \\*
                &                                &  &       & \hfill $\WAIC_{best} - \WAIC_{true}$:        & \textbf{0.0059} &      \\

\midrule
Fr              & Fr                             & -0.0741 & 87\,m & Ga                             & -0.0618 & 9\,m\\*
Cl$^{180}$       & Fr\-Cl$^{180}$                 & -0.1513 &       & In\-Ga\-Gu$^{180}$\-Gu$^{270}$\-Gu$^{0}$\-Gu$^{90}$ & -0.1567 &      \\*
Gu$^{270}$     & Fr\-Cl$^{180}$\-Gu$^{270}$     & -0.1707 &       & In\-Ga\-Cl$^{0}$\-Cl$^{90}$\-Cl$^{180}$\-Cl$^{270}$ & -0.1670 &      \\*
                & Fr\-Cl$^{180}$\-Gu$^{270}$\-Cl$^{90}$ & -0.1708 &       & In\-Ga\-Gu$^{270}$\-Cl$^{180}$\-Cl$^{270}$ & -0.1680 &      \\*
                & Fr\-Cl$^{180}$\-Gu$^{270}$\-Cl$^{90}$\-Gu$^{180}$ & -0.1711 &       & In\-Ga\-Gu$^{270}$\-Cl$^{180}$\-Gu$^{90}$ & -0.1695 &      \\*
                & Fr\-Cl$^{180}$\-Gu$^{270}$\-Cl$^{90}$\-Gu$^{180}$\-Cl$^{0}$ & -0.1710 &       & \textbf{In\-Gu$^{270}$\-Cl$^{180}$} & \textbf{-0.1735} &      \\*
                & \textbf{Gu$^{270}$\-Cl$^{180}$\-Fr} & \textbf{-0.1703} &       &                                &  &      \\*
                &                                &  &       & \hfill $\WAIC_{best} - \WAIC_{true}$:        & \textbf{-0.0011} &      \\

\midrule
Gu$^{0}$     & Gu$^{0}$                       & -0.1695 & 47\,m & Ga                             & -0.1477 & 11\,m\\*
Gu$^{180}$        & Gu$^{0}$\-Cl$^{90}$            & -0.3040 &       & In\-Ga\-Gu$^{180}$\-Gu$^{270}$\-Gu$^{0}$\-Gu$^{90}$ & -0.2986 &      \\*
Cl$^{90}$        & Gu$^{0}$\-Cl$^{90}$\-Gu$^{180}$ & -0.3234 &       & In\-Ga\-Cl$^{0}$\-Cl$^{90}$\-Cl$^{180}$\-Cl$^{270}$ & -0.3033 &      \\*
                & Gu$^{0}$\-Cl$^{90}$\-Gu$^{180}$\-Cl$^{180}$ & -0.3233 &       & Ga\-Gu$^{180}$\-Cl$^{90}$\-Cl$^{180}$ & -0.3054 &      \\*
                & \textbf{Gu$^{180}$\-Cl$^{90}$\-Gu$^{0}$} & \textbf{-0.3234} &       & Ga\-Gu$^{180}$\-Cl$^{90}$\-Gu$^{0}$ & -0.3111 &      \\*
                &                                &  &       & \textbf{Gu$^{180}$\-Cl$^{90}$\-Gu$^{0}$} & \textbf{-0.3113} &      \\

    \bottomrule
  \end{longtable}
  
  \captionof{table}{The model selection histories for 3-element mixtures with constant $\phi$ and variable $\theta$}
  \label{tab:search3T}
  \addtocounter{table}{-1}
  \centering
  \begin{longtable}{p{0.8cm}|p{3.3cm}ll|p{3.45cm}ll}
    \toprule
    \maketabletitle
    \midrule
    \endhead
    Ga           & Fr                             & -0.0177 & 66\,m & Ga                             & -0.0142 & 13\,m\\*
Cl$^{90}$        & Fr\-Gu$^{270}$                 & -0.1284 &       & In\-Ga\-Gu$^{180}$\-Gu$^{270}$\-Gu$^{0}$\-Gu$^{90}$ & -0.1291 &      \\*
Gu$^{0}$          & Fr\-Gu$^{270}$\-Gu$^{0}$       & -0.1407 &       & In\-Ga\-Cl$^{0}$\-Cl$^{90}$\-Cl$^{180}$\-Cl$^{270}$ & -0.1289 &      \\*
                & Fr\-Gu$^{270}$\-Gu$^{0}$\-Cl$^{0}$ & -0.1423 &       & In\-Cl$^{0}$\-Gu$^{270}$\-Gu$^{0}$ & -0.1317 &      \\*
                & Fr\-Gu$^{270}$\-Gu$^{0}$\-Cl$^{0}$\-Cl$^{180}$ & -0.1435 &       & In\-Cl$^{0}$\-Cl$^{90}$\-Gu$^{0}$ & -0.1346 &      \\*
                & Fr\-Gu$^{270}$\-Gu$^{0}$\-Cl$^{0}$\-Cl$^{180}$\-In & -0.1432 &       & In\-Cl$^{0}$\-Cl$^{90}$\-Cl$^{180}$ & -0.1301 &      \\*
                & \textbf{Gu$^{0}$\-Gu$^{270}$}  & \textbf{-0.1451} &       & \textbf{Ga\-Cl$^{90}$\-Cl$^{180}$} & \textbf{-0.1313} &      \\*
                & \hfill $\WAIC_{best} - \WAIC_{true}$:        & \textbf{0.0132} &       & \hfill $\WAIC_{best} - \WAIC_{true}$:        & \textbf{0.0068} &      \\

\midrule
Fr               & Fr                             & -0.0265 & 71\,m & Ga                             & -0.0192 & 9\,m\\*
Cl$^{90}$         & Fr\-Gu$^{270}$                 & -0.1290 &       & In\-Ga\-Gu$^{180}$\-Gu$^{270}$\-Gu$^{0}$\-Gu$^{90}$ & -0.1411 &      \\*
Gu$^{0}$      & Fr\-Gu$^{270}$\-Gu$^{0}$       & -0.1445 &       & In\-Ga\-Cl$^{0}$\-Cl$^{90}$\-Cl$^{180}$\-Cl$^{270}$ & -0.1429 &      \\*
                & Fr\-Gu$^{270}$\-Gu$^{0}$\-Cl$^{180}$ & -0.1450 &       & In\-Ga\-Gu$^{180}$\-Cl$^{90}$\-Cl$^{180}$ & -0.1474 &      \\*
                & Fr\-Gu$^{270}$\-Gu$^{0}$\-Cl$^{180}$\-Cl$^{0}$ & -0.1466 &       & In\-Ga\-Gu$^{180}$\-Cl$^{90}$\-Gu$^{0}$ & -0.1472 &      \\*
                & Fr\-Gu$^{270}$\-Gu$^{0}$\-Cl$^{180}$\-Cl$^{0}$\-In & -0.1468 &       & \textbf{In\-Cl$^{90}$\-Gu$^{0}$} & \textbf{-0.1477} &      \\*
                & \textbf{Gu$^{0}$\-Gu$^{270}$}  & \textbf{-0.1451} &       &                                &  &      \\*
                & \hfill $\WAIC_{best} - \WAIC_{true}$:        & \textbf{0.0109} &       & \hfill $\WAIC_{best} - \WAIC_{true}$:        & \textbf{0.0010} &      \\

\midrule
Fr           & Fr                             & -0.0129 & 61\,m & Ga                             & -0.0185 & 6\,m\\*
Cl$^{180}$        & Fr\-Gu$^{270}$                 & -0.1105 &       & In\-Ga\-Gu$^{180}$\-Gu$^{270}$\-Gu$^{0}$\-Gu$^{90}$ & -0.1309 &      \\*
Gu$^{270}$      & Fr\-Gu$^{270}$\-Gu$^{0}$       & -0.1237 &       & In\-Ga\-Cl$^{0}$\-Cl$^{90}$\-Cl$^{180}$\-Cl$^{270}$ & -0.1326 &      \\*
                & Fr\-Gu$^{270}$\-Gu$^{0}$\-Cl$^{180}$ & -0.1254 &       & In\-Gu$^{270}$\-Cl$^{180}$ & -0.1393 &      \\*
                & Fr\-Gu$^{270}$\-Gu$^{0}$\-Cl$^{180}$\-Gu$^{180}$ & -0.1248 &       & In\-Gu$^{270}$\-Gu$^{0}$ & -0.1334 &      \\*
                & \textbf{Gu$^{0}$\-Gu$^{270}$\-Fr} & \textbf{-0.1234} &       & \textbf{In\-Gu$^{270}$\-Gu$^{0}$} & \textbf{-0.1326} &      \\*
                & \hfill $\WAIC_{best} - \WAIC_{true}$:        & \textbf{0.0094} &       & \hfill $\WAIC_{best} - \WAIC_{true}$:        & \textbf{0.0088} &      \\

\midrule
Gu$^{0}$     & Gu$^{0}$                       & -0.0756 & 55\,m & Ga                             & -0.0454 & 7\,m\\*
Gu$^{180}$        & Gu$^{0}$\-Cl$^{90}$            & -0.2380 &       & In\-Ga\-Gu$^{180}$\-Gu$^{270}$\-Gu$^{0}$\-Gu$^{90}$ & -0.2476 &      \\*
Cl$^{90}$        & Gu$^{0}$\-Cl$^{90}$\-Cl$^{0}$  & -0.2556 &       & In\-Ga\-Cl$^{0}$\-Cl$^{90}$\-Cl$^{180}$\-Cl$^{270}$ & -0.2459 &      \\*
                & Gu$^{0}$\-Cl$^{90}$\-Cl$^{0}$\-Ga & -0.2591 &       & Ga\-Cl$^{0}$\-Gu$^{270}$\-Gu$^{0}$ & -0.2493 &      \\*
                & Gu$^{0}$\-Cl$^{90}$\-Cl$^{0}$\-Ga\-Cl$^{270}$ & -0.2590 &       & Ga\-Cl$^{0}$\-Cl$^{90}$\-Gu$^{0}$ & -0.2559 &      \\*
                & \textbf{Cl$^{0}$\-Cl$^{90}$\-Gu$^{0}$} & \textbf{-0.2555} &       & \textbf{Cl$^{0}$\-Cl$^{90}$\-Gu$^{0}$} & \textbf{-0.2538} &      \\*
                & \hfill $\WAIC_{best} - \WAIC_{true}$:        & \textbf{0.0026} &       & \hfill $\WAIC_{best} - \WAIC_{true}$:        & \textbf{0.0006} &      \\

    \bottomrule
  \end{longtable}
  
%  \end{table}

\end{document}